\documentclass[journal,12pt,onecolumn,draftclsnofoot]{IEEEtran}

\usepackage{color}
\usepackage{multirow}
\usepackage{graphicx}
\usepackage{epstopdf}
\usepackage[cmex10]{amsmath}
\usepackage{bm} 
\usepackage{amssymb}

\usepackage{multirow}
\usepackage{amsthm}
\usepackage{cite}
\usepackage{balance}
\usepackage{acronym} 
\usepackage{algorithm, algcompatible}

\newcommand{\diag}{\mathop{\rm diag}}

\DeclareMathOperator*{\argmax}{\arg\!\max}

\algnewcommand\INPUT{\item[\textbf{Input:}]}
\algnewcommand\OUTPUT{\item[\textbf{Output:}]}

\allowdisplaybreaks

\begin{document}

\title{Zero Forcing Uplink Detection through Large-Scale RIS: System Performance and Phase Shift Design} 
\author{Nikolaos~I.~Miridakis,~\IEEEmembership{Senior Member,~IEEE}, Theodoros~A.~Tsiftsis,~\IEEEmembership{Senior Member,~IEEE} and Rugui~Yao,~\IEEEmembership{Senior Member,~IEEE}
\thanks{N.~I.~Miridakis is with the Department of Informatics and Computer Engineering, University of West Attica, Aegaleo 12243, Greece (e-mail:~nikozm@uniwa.gr).}
\thanks{T.~A.~Tsiftsis is with the School of Intelligent Systems Science \& Engineering, Jinan University, Zhuhai Campus, Zhuhai 519070, China, and also with the University of Thessaly, Greece. (e-mail:~tsiftsis@ieee.org).}
\thanks{R.~Yao is with the School of Electronics and Information, Northwestern Polytechnical University, Xi'an 710072, China  (e-mail:~yaorg@nwpu.edu.cn).}
}


\maketitle

\begin{abstract}
A multiple-input multiple-output wireless communication system is analytically studied, which operates with the aid of a large-scale reconfigurable intelligent surface (LRIS). LRIS is equipped with multiple passive elements with discrete phase adjustment capabilities, and independent Rician fading conditions are assumed for both the transmitter-to-LRIS and LRIS-to-receiver links. A direct transceiver link is also considered which is modeled by Rayleigh fading distribution. The system performance is analytically studied when the linear yet efficient zero-forcing detection is implemented at the receiver. In particular, the outage performance is derived in closed-form expression for different system configuration setups with regards to the available channel state information (CSI) at the receiver. In fact, the case of both perfect and imperfect CSI is analyzed. Also, an efficient phase shift design approach at LRIS is introduced, which is linear on the number of passive elements and receive antennas. The proposed phase shift design can be applied on two different modes of operation; namely, when the system strives to adapt either on the instantaneous or statistical CSI. Finally, some impactful engineering insights are provided, such as how the channel fading conditions, CSI, discrete phase shift resolution, and volume of antenna/LRIS element arrays impact on the overall system performance. 
\end{abstract}


\IEEEpeerreviewmaketitle

\section{Introduction}
\IEEEPARstart{W}{ith} the rapid growth of wireless communication technology, research and development for beyond 5G and even 6G networking infrastructures has already been initiated worldwide \cite{3gppR18,j:whitepaper2020}. Some of the most important key performance indicators (KPIs) include high energy efficiency, increased spectral efficiency, increased density, enhanced coverage and service availability, as well as increased area traffic capacity \cite{j:TariqFaisal2020}. Reconfigurable intelligent surfaces (RIS), also known as intelligent reflecting metasurfaces, which are emerging at the forefront of wireless communications research nowadays represent a prominent candidate to facilitate the support of the said KPIs \cite{j:ChengLeiDai2022,j:SwindlehurstZhou2022}.

To further boost the system performance, multiple input-multiple output (MIMO) signal transmission can be used in conjunction with RIS \cite{j:DiBoya2020,j:WuQingqing2020,j:NadeemQurrat2020,j:ZhiKangda2021,j:XiaoGechuan2022,j:ZhiKanda2022}. Nevertheless, the main challenge in these systems is the phase shift adaptation/optimization. To this end, channel state information (CSI) of the associated links is unfortunately a prerequisite which in turn burdens the overall computational complexity proportionally to the number of active antennas and passive RIS elements {\color{black}\cite{j:WeiXiuhong2021,j:KimSucheol2022,j:WeiLiHuangSha2022,j:GuoYabo2022,j:WeiLiHuangChongwen2022,j:WeiLiHuang2022}}.

\subsection{Related State of the Art}
Beamforming optimization design for RIS-assisted multiple input-single output (MISO) systems was studied in \cite{j:DiBoya2020} and \cite{j:WuQingqing2020}, under discrete phase shifts. Cascaded line-of-sight (LoS) signal propagation and Rayleigh channel fading links were respectively considered in \cite{j:DiBoya2020} and \cite{j:WuQingqing2020}; yet assuming perfect (ideal) CSI conditions. 

The realistic estimated (i.e., imperfect) CSI case was considered in \cite{j:NadeemQurrat2020} for hybrid Rayleigh channel faded links and pure LoS signal propagation, yet assuming a continuous (ideal) phase shift design. The performance of a massive MIMO RIS-enabled system under Rician channel fading and Rayleigh faded direct links was analytically studied by the authors of \cite{j:ZhiKangda2021}. Therein, however, perfect CSI was assumed, while a heuristic algorithm was presented to optimize the continuous RIS phase shifts based on a statistical CSI knowledge. 

In \cite{j:XiaoGechuan2022} and \cite{j:ZhiKanda2022}, the performance of a RIS-assisted MIMO system was studied, when one of the two RIS-enabled links is modeled as Rician fading channels while the other is described by pure LoS conditions. Also, a phase shift design was proposed in these works based on a statistical CSI. Nevertheless, the pure LoS channel fading condition is a relaxed assumption to facilitate the analysis and may not sufficiently correspond to various practical illustrations. Notably, whenever the RIS-user distance is relatively small, the pure LoS assumption may not be enough to accurately define the total channel fading. This occurs because multipath propagation is always present, especially in dense outdoor terrestrials and/or close-distant indoor environments. {\color{black}In \cite{j:ZhangLiuJun2021}, the performance of a RIS-assisted MIMO system was analytically studied and optimized when all the included links undergo Rician channel fading. Therein, however, asymptotically high antenna arrays were considered for all the nodes while a continuous phase shift design at RIS was assumed.} 

\subsection{Motivation and Contribution}
Motivated by the aforementioned observations, a MIMO wireless communication system operating through a large-scale RIS (LRIS) is analytically studied into this paper. In particular, the spatial multiplexing mode of MIMO operation is considered where multiple signals are simultaneously transmitted and received by utilizing the linear yet effective zero forcing (ZF) detection. The assumption of LRIS is reasonable since the corresponding arrays typically consist of passive elements; thereby a large-scale (yet finite) array with low-cost passive elements becomes practically feasible. Moreover, all the LRIS-enabled links are modeled by independent Rician channel fading while Rayleigh fading is considered for the transceiver's direct links. In addition, discrete-only phase shifts occur at LRIS, which is a realistic condition. 

Unlike the current state-of-the-art, in this paper, channel fading is present at all the available channel links, whereas the phase shifts at LRIS are discrete; thereby they introduce an unavoidable phase mismatch in comparison to an ideal continuous phase shift adaptation. The contributions of this work are summarized as follows:
\begin{itemize}
	\item New and rather simple closed-form expressions regarding the key statistics of the received signal-to-noise ratio (SNR) are obtained. The introduced analysis is valid for the perfect CSI case (ideal, it may serve as a performance benchmark), whereas it also considers the realistic case of imperfect/estimated CSI acquisition at the receiver.
	\item Based on the latter derivations, new engineering insights are emerged, such as how the interrelation between the number of quantization bits at the discrete LRIS phase shifts, channel fading conditions and transceiver array volume affect the system performance.
	\item A new phase shift design is formulated in a straightforward closed-form expression, which produces a linear computational complexity with respect to the number of LRIS elements and receive antennas. According to this design, the minimum received SNR is being maximized up to a certain convergence level. Specifically, two distinct algorithms are devised, which operate according to an instantaneous or statistical CSI feedback adaptation at LRIS, correspondingly.   
\end{itemize}
  
\subsection{Organization of the paper}	
The rest of this paper is organized as follows. This Section continues with some notational definitions for the most important mathematical symbols used in the subsequent analysis. In Section II, the considered system and signal models are described in detail. Key statistical derivations and corresponding performance measures for the case of perfect and imperfect CSI are obtained in closed form in Section III. In Section IV, the proposed phase shift design at LRIS is explicitly defined for both the cases when LRIS follows instantaneous CSI changes or statistical ones. In Section V, the proposed framework is validated and cross-compared with simulation results, while some useful engineering insights are revealed. Finally, Section VI concludes the paper.

{\it Notation:} Vectors and matrices are represented by lowercase and uppercase bold typeface letters, respectively, whereas $\mathbf{I}_{v}$ stands for the $v\times v$ identity matrix and $\mathbf{0}_{v}$ is the $v\times v$ null matrix. A diagonal matrix with entries $x_{1},\cdots,x_{n}$ is defined as $\diag\{x_{i}\}^{n}_{i=1}$, $\mathbf{X}^{-1}$ is the inverse of $\mathbf{X}$ and $\mathbf{X}^{\dagger}$ is the Moore-Penrose pseudoinverse of $\mathbf{X}$. $[\mathbf{X}]_{k,l}$ stands for the entry at the $k^{\rm th}$ row and $l^{\rm th}$ column of $\mathbf{X}$, $[\mathbf{X}]_{k,:}$ is the $k^{\rm th}$ row vector of $\mathbf{X}$ and $[\mathbf{X}]_{:,l}$ is the $l^{\rm th}$ column vector of $\mathbf{X}$. Superscripts $(\cdot)^{\star}$, $(\cdot)^{\mathcal{T}}$ and $(\cdot)^{\mathcal{H}}$ denote scalar conjugate transpose, transpose and vector/matrix conjugate (Hermitian) transpose, respectively; $\det[\cdot]$ and ${\rm Tr}[\cdot]$ represent the determinant and trace of a given matrix, respectively; $\otimes$ is the Kronecker product, $|\cdot|$ represents absolute (scalar) value, ${\rm Re}\{\cdot\}$ is the real part of a complex value, $\angle[\cdot]$ is the phase of a complex argument, and $j\triangleq \sqrt{-1}$. $\mathbb{E}[\cdot]$ is the expectation operator, $\mathbb{E}[\mathbf{X}|\mathbf{Y}]$ is the conditional expectation of $\mathbf{X}$ given $\mathbf{Y}$, symbol $\overset{\text{d}}=$ means equality in distribution and $\overset{\text{d}}\approx$ defines almost sure convergence (asymptotically) in distribution. $f_{X}(\cdot)$ and $F_{X}(\cdot)$ represent the probability density function and cumulative distribution function (CDF) of a random variable (RV) $X$, respectively. $\mathcal{CN}(\mu,\sigma^{2})$ defines a complex-valued Gaussian RV with mean $\mu$ and variance $\sigma^{2}$. For a $N\times M$ $\mathbf{X}\overset{\text{d}}=\mathcal{CN}(\mathbf{A},\mathbf{B})$, a complex-valued central and non-central Wishart $M\times M$ matrix is defined as $\mathbf{X}^{\mathcal{H}}\mathbf{X}\overset{\text{d}}=\mathcal{W}_{M}(N,\mathbf{\Sigma})$ (only when $\mathbf{A}=\mathbf{0}_{M}$) and $\mathcal{W}_{M}(N,\mathbf{\Omega},\mathbf{\Sigma})$, respectively, with $N$ degrees-of-freedom (DoF), covariance matrix $\mathbf{\Sigma}$, and non-centrality (positive definite) matrix $\mathbf{\Omega}=\mathbf{B}^{-1}\mathbf{A}^{\mathcal{H}}\mathbf{A}$ \cite{b:multivariate}. Also, $\Gamma(\cdot)$ denotes the Gamma function \cite[Eq. (8.310.1)]{tables}, $\Gamma(\cdot,\cdot)$ denotes the upper incomplete Gamma function \cite[Eq. (8.350.2)]{tables}, ${}_1F_{1}(\cdot,\cdot;\cdot)$ is the Kummer's confluent hypergeometric function \cite[Eq. (9.210.1)]{tables}, and ${\rm sinc}(x)=\sin(x)/x$ is the sinc function. Finally, $\mathcal{O}(\cdot)$ represents the Landau symbol; i.e., for two arbitrary functions $f(x)$ and $g(x)$, it holds that $f(x)=\mathcal{O}(g(x))$ when $|f(x)|\leq v |g(x)|\: \forall x\geq x_{0},\{v,x_{0}\}\in \mathbb{R}$.

\section{System and Signal Model}
Consider a wireless communication system with $M$ (co-located) transmit and $N\geq M$ receive antennas operating over a quasi-static block-fading channel. A practical application may correspond to the uplink transmission when the transmit and receive nodes are the system user and base station, respectively. The end-to-end communication is assisted by an intermediate LRIS equipped with $L$ passive elements. It is also assumed that both the size of each element and the inter-element spacing are equal to half of the signal wavelength; such that the associated channels undergo independent fading \cite{j:RenzoZappone20}.\footnote{According to \cite{j:EmilLuca2021}, channel fading (spatial) correlation is always present in practical RIS illustrations. However, it was recently proved in \cite[Prop.~6 and Fig.~7]{j:WangBadiu2022} that, for the large-scale RIS case, the effect of spatial correlation introduces negligible impact to the system performance compared to the spatially independent assumption. Thus, to facilitate the following analysis, independent channel-faded links are considered hereinafter.} It is assumed that the transmitter-to-LRIS and LRIS-to-receiver links undergo independent Rician channel fading due to their relatively close distance and the (potential) presence of a strong LoS channel gain component. On the other hand, the transceiver links are subject to independent Rayleigh channel fading, due to a relatively high link distance and the presence of intense signal attenuation in a rich scattering environment. The spatial multiplexing mode of operation is utilized, where all the given streams are simultaneously transmitted and received via ZF detection. 

\begin{figure}[!t]
\centering
\includegraphics[trim=.5cm .5cm .5cm 0.0cm, clip=true,totalheight=0.27\textheight]{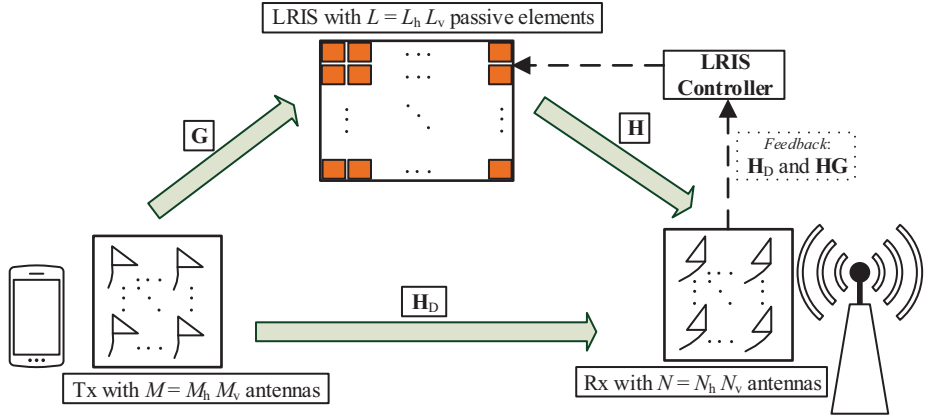}
\caption{The considered system model.}
\label{fig1}
\end{figure}

More specifically, the received signal reads as 
\begin{align}
\nonumber
\mathbf{y}&=\sqrt{p}\mathbf{H}_{\rm tot}\mathbf{s}+\mathbf{n}\\
&=\sqrt{p}\left(\mathbf{H}_{\rm D}+\mathbf{H}\mathbf{\Phi}\mathbf{G}\right)\mathbf{s}+\mathbf{n},
\label{received}
\end{align}
where $\mathbf{y} \in \mathbb{C}^{N \times 1}$, $p$ is the transmit SNR per antenna, $\mathbf{s}\in \mathbb{C}^{M \times 1}$ is the transmit signal and
\begin{align}
\mathbf{H}_{\rm tot}\triangleq \mathbf{H}_{\rm D}+\mathbf{H}\mathbf{\Phi}\mathbf{G}.
\label{channelTotal}
\end{align}
Also, $\mathbf{H}_{\rm D}\in \mathbb{C}^{N \times M}$ denotes the channel matrix between the transceiver direct links; $\mathbf{H}\in \mathbb{C}^{N \times L}$ and $\mathbf{G}\in \mathbb{C}^{L \times M}$ are the channel matrices of the LRIS-to-receiver and transmitter-to-LRIS links, respectively; $\mathbf{\Phi}\triangleq \diag\{e^{j \phi_{i}}\}^{L}_{i=1}$ denotes the phase rotations at LRIS; and $\mathbf{n}\in \mathbb{C}^{N \times 1}$ defines the additive white Gaussian noise (AWGN) at the receiver. In practice, $\{\phi_{i}\}^{L}_{i=1}$ can only be configured from a given discrete-phase set $\mathcal{S}$, where $\mathcal{S}\in[-\pi,\pi]$; such that the cardinality of $\mathcal{S}$ is denoted by $|\mathcal{S}|=2^{q}$ with a $q-$bit quantization satisfying $q\geq 1$. 

Further, it holds that $\mathbf{H}_{\rm D}\overset{\text{d}}=\mathcal{CN}(\mathbf{0},\beta_{\rm UB}\mathbf{I}_{N})$ with $\beta_{\rm UB}$ standing for the (known) large-scale channel gain of the transceiver link. In a similar basis, $\mathbf{H}\overset{\text{d}}=\mathcal{CN}(\mathbf{M}_{\rm LB},\beta_{\rm LB}\mathbf{I}_{N})$, where $\mathbf{M}_{\rm LB}$ is the rank-one LoS (mean) channel fading matrix of the LRIS-to-receiver link. Also, $\mathbf{G}\overset{\text{d}}=\mathcal{CN}(\mathbf{M}_{\rm UL},\beta_{\rm UL}\mathbf{I}_{L})$, where $\mathbf{M}_{\rm UL}$ is the rank-one LoS channel fading matrix of the transmitter-to-LRIS link. Without loss of generality, we consider a uniform planar array (UPA) structure for both LRIS and transceiver nodes so as to meet practical illustration. Thereby, it holds that \cite{j:KhiongYong2005} 
\begin{align}
\nonumber
\mathbf{M}_{\rm LB}=&\mathbf{a}_{\rm LRIS-Rx}\left[\theta^{\rm rx}_{\rm LB};\psi^{\rm rx}_{\rm LB},\theta^{\rm rx}_{\rm LB}\right]\\
&\times \mathbf{a}^{\mathcal{H}}_{\rm Rx-LRIS}\left[\theta^{\rm tx}_{\rm LB};\psi^{\rm tx}_{\rm LB},\theta^{\rm tx}_{\rm LB}\right],
\label{LOSarrays}
\end{align}
where $\mathbf{a}_{\rm LRIS-Rx}[\cdot;\cdot,\cdot]$ denotes the steering vector at LRIS regarding the LRIS-receiver link and $\mathbf{a}_{\rm Rx-LRIS}[\cdot,\cdot]$ denotes the corresponding steering vector at the receiver. Also, $\{\theta^{\rm rx}_{\rm LB}, \psi^{\rm rx}_{\rm LB}\}$ and $\{\theta^{\rm tx}_{\rm LB}, \psi^{\rm tx}_{\rm LB}\}$ represent the vertical and horizontal arrival angles at the receiver and the departure angles at LRIS, respectively. The general form of the aforementioned steering vector between the $i^{\rm th}$ and $j^{\rm th}$ node reads as
\begin{align}
\nonumber
\mathbf{a}_{i-j}[\theta;\theta,\psi]\triangleq &\frac{1}{\sqrt{F_{i}}}\left[1,\ldots,e^{j (F_{\rm v}-1) \pi \sin(\theta)}\right]^{\mathcal{T}}\\
&\otimes \left[1,\ldots,e^{j (F_{\rm h}-1) \pi \sin(\psi)\cos(\theta)}\right]^{\mathcal{T}},
\label{steeringVector}
\end{align}
where $F_{i}=F_{\rm v}F_{\rm h}$ is the length of the latter vector (with respect to the $i^{\rm th}$ node) and $\{F_{\rm v},F_{\rm h}\}$ stand for the number of vertical and horizontal LRIS elements or antennas, respectively. Similarly, we get
\begin{align}
\nonumber
\mathbf{M}_{\rm UL}=&\mathbf{a}_{\rm LRIS-Tx}\left[\theta^{\rm rx}_{\rm LU};\psi^{\rm rx}_{\rm LU},\theta^{\rm rx}_{\rm LU}\right]\\
&\times \mathbf{a}^{\mathcal{H}}_{\rm Tx-LRIS}\left[\theta^{\rm tx}_{\rm LU};\psi^{\rm tx}_{\rm LU},\theta^{\rm tx}_{\rm LU}\right],
\label{LOSarrays2}
\end{align}
where $\mathbf{a}_{\rm LRIS-Tx}[\cdot;\cdot,\cdot]$ denotes the steering vector at LRIS regarding the LRIS-transmitter link and $\mathbf{a}_{\rm Tx-LRIS}[\cdot,\cdot]$ denotes the corresponding steering vector at the transmitter (i.e., user). Also, $\{\theta^{\rm rx}_{\rm LU}, \psi^{\rm rx}_{\rm LU}\}$ and $\{\theta^{\rm tx}_{\rm LU}, \psi^{\rm tx}_{\rm LU}\}$ represent the vertical and horizontal arrival angles at the transmitter and the departure angles at LRIS, respectively. The steering vectors in \eqref{LOSarrays2} are formed as in \eqref{steeringVector}, by setting the appropriate index substitutions.
 
Moreover, the gains $\{\beta_{\rm UB},\beta_{\rm UL},\beta_{\rm LB}\}$ incorporate signal propagation attenuation, antenna gains and shadowing losses, which are assumed either fixed or perfectly known at the receiver.\footnote{Note that the large-scale channel gains are identical for the same link since the inter-element distance at LRIS (or the inter-antenna distance at the transceiver) is considered negligible compared to the corresponding LRIS-to-transceiver distances.} Lastly, it is assumed that $\mathbb{E}[\mathbf{s}\mathbf{s}^{\mathcal{H}}]=\mathbf{I}_{M}$ and $\mathbf{n}\overset{\text{d}}=\mathcal{CN}(\mathbf{0},\mathbf{I}_{N})$. The considered system model is sketched in Fig.~\ref{fig1}.

\section{System Performance}
\subsection{Perfect CSI}
Upon the signal reception, the ZF filter matrix $\mathbf{W}\in \mathbb{C}^{M \times N}$ is applied, yielding to the post-processed signal $\mathbf{r}\triangleq \mathbf{W}\mathbf{y}=(\mathbf{H}_{\rm D}+\mathbf{H}\mathbf{\Phi}\mathbf{G})^{\dagger}\mathbf{y}$. The resultant received SNR is thus given by\footnote{{\color{black}The last equality of \eqref{snr} follows by invoking the property $(\mathbf{H}^{\mathcal{H}}_{\rm tot}\mathbf{H}_{\rm tot})^{-1}=(\det[\mathbf{H}^{\mathcal{H}}_{\rm tot}\mathbf{H}_{\rm tot}])^{-1}\rm{adj}[\mathbf{H}^{\mathcal{H}}_{\rm tot}\mathbf{H}_{\rm tot}]$, where $\rm{adj}[\cdot]$ denotes the adjugate matrix operator. Then, after some straightforward manipulations, the diagonal entries of $\rm{adj}[\mathbf{H}^{\mathcal{H}}_{\rm tot}\mathbf{H}_{\rm tot}]$ can be expressed as $\det[\mathbf{H}^{\mathcal{H}}_{{\rm tot},i}\mathbf{H}_{{\rm tot},i}]$.}}
\begin{align}
\nonumber
\gamma_{i}&=\frac{p}{\left[\left(\mathbf{H}^{\mathcal{H}}_{\rm tot}\mathbf{H}_{\rm tot}\right)^{-1}\right]_{i,i}},\quad 1\leq i\leq M,\\
&=p\frac{\det\left[\mathbf{H}^{\mathcal{H}}_{\rm tot}\mathbf{H}_{\rm tot}\right]}{\det\left[\mathbf{H}^{\mathcal{H}}_{{\rm tot},i}\mathbf{H}_{{\rm tot},i}\right]},
\label{snr}
\end{align}
where $\mathbf{H}_{{\rm tot},i}$ is the deflated version of $\mathbf{H}_{\rm tot}$ with its $i^{\rm th}$ column being removed. Regardless of the perfect CSI condition at the receiver side, there is always a mismatch condition at LRIS due to the presence of discrete-only phase adjustments. To this end, each channel entry of the LRIS-transceiver link can be modeled by 
\begin{align}
\left[\mathbf{H}\mathbf{\Phi}\mathbf{G}\right]_{i,j}=\sum^{L}_{l=1}\left|\left[\mathbf{H}\right]_{i,l}\right|\cdot \left|\left[\mathbf{G}\right]_{l,j}\right|e^{j\omega_{l}}, 
\end{align}
where $\omega_{l}$ defines the deviation of $\phi_{l}$ from the ideal setting and is uniformly distributed over $[-2^{-q}\pi,2^{-q}\pi]$. Then, as $L\rightarrow \infty$, we reach at \cite[Eq. (11)]{j:Badiu2020}
\begin{align}
f_{\left|\left[\mathbf{H}\mathbf{\Phi}\mathbf{G}\right]_{i,j}\right|}(x;m,\bar{\gamma})\approx \frac{2 m^{m}}{\Gamma(m)\bar{\gamma}^{m}}x^{2 m-1}\exp\left(-\frac{m x^{2}}{\bar{\gamma}}\right),
\label{pdfenvelope}
\end{align} 
where 
\begin{align} 
m=\frac{L \xi^{2}_{1} \alpha^{4}}{\left[2(1+\xi_{2}-2 \xi^{2}_{1} \alpha^{4})\right]}\textrm{ and } \bar{\gamma}=\xi^{2}_{1}\alpha^{4} L^{2}, 
\label{params}
\end{align} 
while 
\begin{align} 
\xi_{1}={\rm sinc}(2^{-q}\pi);\ \xi_{2}={\rm sinc}(2^{-q+1}\pi) \textrm{ and } \alpha=\sqrt{a_{\mathbf{H}}a_{\mathbf{G}}}. 
\label{xidef}
\end{align}
{\color{black}Also, $a_{j}$ denotes the average fading gain of the $j^{\rm th}$ channel with $j=\{\mathbf{H},\mathbf{G}\}$. Specifically, $a_{\mathbf{H}}\triangleq \mathbb{E}[|[\mathbf{H}]_{u,v}|^{2}]\forall u,v$ and $a_{\mathbf{G}}\triangleq \mathbb{E}[|[\mathbf{G}]_{r,p}|^{2}]\forall r,p$.} For Rician fading, $a_{j}$ stems as 
\begin{align}
a_{j}=\sqrt{\frac{\pi}{4(\kappa_{j}+1)}}{}_1F_{1}(-1/2,1;-\kappa_{j}), 
\label{alphapar}
\end{align}
where $\kappa_{j}$ is the Rician $K-$factor of the $j^{\rm th}$ channel link. Notably, \eqref{pdfenvelope} implies that the envelope of the cascaded LRIS-included received signal with reflected phase imperfections approaches the Nakagami-$m$ distribution. It is well-known that Nakagami-$m$ distribution closely resembles Rician distribution with a modified Rician $K-$factor, such that $f_{\left|\left[\mathbf{H}\mathbf{\Phi}\mathbf{G}\right]_{i,j}\right|}(x;m,\bar{\gamma})\approx f_{\rm Rice}(x;K,\bar{\gamma})$ where\cite[(Eq. 2.26)]{b:AlouiniSimon}\footnote{Recall that $K$ denotes the Rician factor of the equivalent (virtual) cascaded Tx-LRIS-Rx channel, whereas $\kappa_{i}$ is the Rician factor of the individual $i^{\rm th}$ channel.} 
\begin{align}
K\triangleq \sqrt{m^{2}-m}+m-1. 
\end{align}
The latter parameter is valid when $m\geq 1$, which is met for the considered case with large $L$ and $q\geq 1$. Doing so, the equivalent cascaded channel can be approximately remodeled as 
\begin{align}
\mathbf{H}\mathbf{\Phi}\mathbf{G}\approx \sqrt{\beta_{\rm LB}\beta_{\rm UL}}\left[\sqrt{\frac{K}{K+1}}\mathbf{X}+\sqrt{\frac{1}{K+1}}\mathbf{Y}\right],
\label{CascadedChannelStructure}
\end{align}
where 
\begin{align}
\mathbf{X}=\mathbf{M}_{\rm LB}\mathbf{\Phi}\mathbf{M}_{\rm UL}
\label{meanchannel}
\end{align}
is the deterministic (mean) LoS channel, which is still a rank-one matrix satisfying the properties back in \eqref{LOSarrays} and \eqref{LOSarrays2}. The random part $\mathbf{Y}$ is a zero-mean unit-variance complex-valued Gaussian matrix which is full-rank and denotes the non-LoS scattered components. A proper normalization has been performed at $\mathbf{X}$ (i.e., $\mathbf{X}\mapsto \sqrt{N M}\mathbf{X}/\|\mathbf{X}\|$, where $\|\mathbf{X}\|$ denotes the Frobenius norm of $\mathbf{X}$), such that $\|\mathbf{X}\|^{2}=\mathbb{E}[\|\mathbf{Y}\|^{2}]=N M$, in order to preserve the power ratio
\begin{align}
\frac{\frac{K}{K+1}\|\mathbf{X}\|^{2}}{\frac{1}{K+1}\mathbb{E}[\|\mathbf{Y}\|^{2}]}=K.
\label{powerRatio}
\end{align}
Hence, according to the channel structure as per \eqref{channelTotal} and \eqref{CascadedChannelStructure}, it can be shown that 
\begin{align}
\nonumber
\mathbf{H}^{\mathcal{H}}_{\rm tot}\mathbf{H}_{\rm tot}&\overset{\text{d}}\approx\mathcal{W}_{M}\left(\scriptstyle N;\frac{K \beta_{\rm LB}\beta_{\rm UL}}{(K+1)\beta_{\rm UB}+\bar{\gamma}\beta_{\rm LB}\beta_{\rm UL}} \mathbf{X}^{\mathcal{H}}\mathbf{X};\left[\beta_{\rm UB}+\frac{\bar{\gamma}\beta_{\rm LB}\beta_{\rm UL}}{K+1}\right]\mathbf{I}_{M}\right)\\
&\approx \mathcal{W}_{M}(N;\mathbf{\Sigma}),
\label{wishart}
\end{align}
where 
\begin{align}
\mathbf{\Sigma}\triangleq \left(\beta_{\rm UB}+\frac{\bar{\gamma}\beta_{\rm LB}\beta_{\rm UL}}{K+1}\right)\mathbf{I}_{M}+\frac{K \beta_{\rm LB}\beta_{\rm UL}}{N (K+1)}\mathbf{X}^{\mathcal{H}}\mathbf{X}.
\label{sigma}
\end{align}
The latter expression of \eqref{wishart} turns the non-central Wishart matrix into an approximate central Wishart one with a modified covariance matrix, as firstly suggested in \cite[\S~4]{j:steyn1972approximations}. The effectiveness of the said approximation has been tested and verified in numerous relevant works (e.g., see \cite{j:MatthaiouMckaySmith2010,j:SiriteanuMiyanaga2012} and references therein). Therefore, based on \eqref{wishart}, the received SNR in \eqref{snr} is approached by a chi-squared RV with CDF given by
\begin{align}
F_{\gamma_{i}}(x)\approx 1-\exp\left(-\frac{x [\mathbf{\Sigma}^{-1}]_{i,i}}{p}\right)\sum^{N-M}_{k=0}\frac{\left(x [\mathbf{\Sigma}^{-1}]_{i,i}/p\right)^{k}}{k!}.
\label{CDFSNR}
\end{align}
{\color{black}To verify the accuracy of the proposed approach, Fig.~\ref{fig11} cross-compares \eqref{CDFSNR} and the empirical CDF (with $95\%$ confidence level) obtained by simulating the resultant SNR utilizing \eqref{channelTotal} and \eqref{snr} for different $L$ values. As expected, the approach gets tighter for increasing RIS elements.

\begin{figure}[!t]
\centering
\includegraphics[trim=2cm .5cm .5cm 0.0cm, clip=true,totalheight=0.4\textheight]{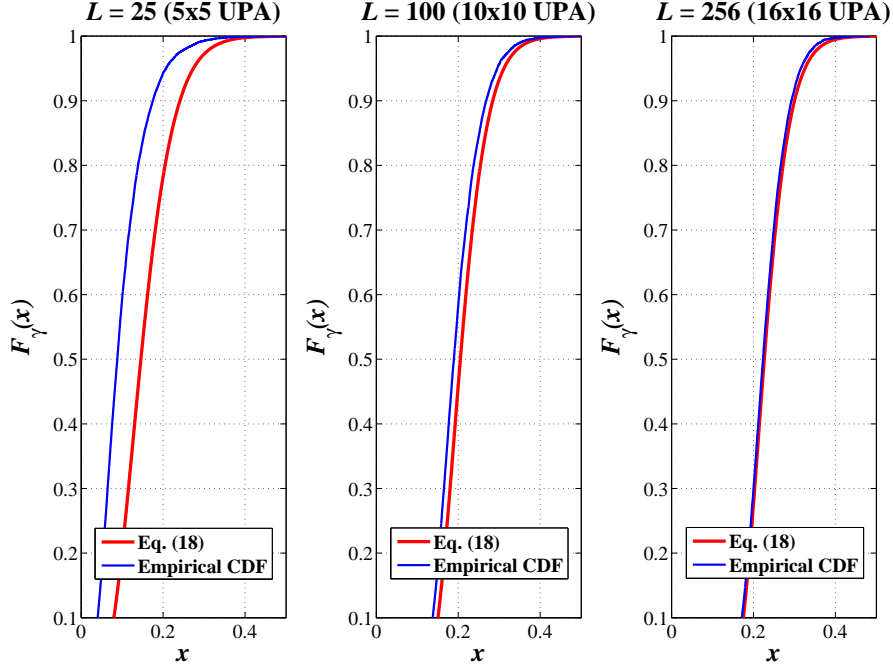}
\caption{Cross-comparison of the derived results with simulation empirical data. Without loss of generality and for ease of clarity, we set $p=0$dB, $q=2$, $\beta_{\rm UB}=\beta_{\rm LB}=\beta_{\rm UL}=10^{-1}$, $M=4$, $N=8$, and $\kappa_{\mathbf{H}}=\kappa_{\mathbf{G}}\triangleq 10$dB.}
\label{fig11}
\end{figure}}

We proceed by reformulating the rank-one matrix $\mathbf{X}$ as $\mathbf{X}\triangleq \mathbf{a}\mathbf{b}^{\mathcal{H}}$. Then, it holds that $\|\mathbf{a}\|=1$ and $\|\mathbf{b}\|=\sqrt{N M K/(K+1)}$ to ensure that the Rician $K-$factor results to the correct relationship with the underlying channel fading as per \eqref{powerRatio}. Therefore, applying the Sherman-Morrison formula to $\mathbf{\Sigma}^{-1}$ and after some straightforward manipulations, we get
\begin{align}
\nonumber
[\mathbf{\Sigma}^{-1}]_{i,i}&=\frac{1}{\beta_{\rm UB}+\frac{\bar{\gamma}\beta_{\rm LB}\beta_{\rm UL}}{K+1}}\\
&-\frac{\frac{K^{2}\beta_{\rm LB}\beta_{\rm UL}}{(K+1)^{2}}}{\left(\beta_{\rm UB}+\frac{\bar{\gamma}\beta_{\rm LB}\beta_{\rm UL}}{K+1}\right)^{2}\left(1+\frac{\beta_{\rm LB}\beta_{\rm UL}K M \left(\frac{K}{K+1}\right)}{\beta_{\rm UB}+\beta_{\rm LB}\beta_{\rm UL}\bar{\gamma}+\beta_{\rm UB} K}\right)},
\label{sigmainverse}
\end{align} 
for $1\leq i \leq M$. 

It is noteworthy that $K\propto L$ and $\bar{\gamma}\propto L^{2}$, according to \eqref{params}. As $L\rightarrow \infty$, it can be easily shown that the rightmost part of \eqref{sigmainverse} tends to zero, yielding
\begin{align}
\nonumber
[\mathbf{\Sigma}^{-1}]_{i,i}&\approx \left(\beta_{\rm UB}+\frac{\bar{\gamma}\beta_{\rm LB}\beta_{\rm UL}}{K+1}\right)^{-1},\quad L\rightarrow \infty,\\
&=\left[\beta_{\rm UB}+\beta_{\rm LB}\beta_{\rm UL} L\left(1-2 \xi^{2}_{1} \alpha^{4}+\xi_{2}\right)\right]^{-1}.
\label{sigmainverseAsymptotic}
\end{align}
{\color{black}Based on the latter result, the CDF of the received SNR in \eqref{CDFSNR} can asymptotically approached by 
\begin{align}
\nonumber
F_{\gamma_{i}}(x)&\approx \exp\left(-\frac{x [\mathbf{\Sigma}^{-1}]_{i,i}}{p}\right)\sum^{\infty}_{k=N-M+1}\frac{\left(x [\mathbf{\Sigma}^{-1}]_{i,i}/p\right)^{k}}{k!}\\
\nonumber
&\approx \frac{1}{(N-M+1)!}\\
&\times \left(\frac{x/p}{\beta_{\rm UB}+\beta_{\rm LB}\beta_{\rm UL} L\left(1-2 \xi^{2}_{1} \alpha^{4}+\xi_{2}\right)}\right)^{N-M+1}, \quad L\rightarrow \infty,
\label{CDFSNRasy}
\end{align}
which follows by turning \eqref{CDFSNR} into an infinite series representation, while observing that the exponential term tends to unity and the lowest sum term becomes dominant as $L\rightarrow \infty$.}

The asymptotic CDF expression in \eqref{CDFSNRasy} reveals that the diversity order of the considered system is not affected by the presence of LRIS (and, most insightfully, by its discrete phase mismatches), while the order of $L$ influences only the coding/array gain. Also, \eqref{CDFSNRasy} is valid for an arbitrary range of the transmit SNR $p$. For the extreme cases of $q=1$ (lowest resolution) and $q\rightarrow \infty$ (highest resolution), the aforementioned expressions can be further relaxed. Specifically, by referring back to \eqref{xidef}, it can be shown that $\xi_{1}=2/\pi$ and $\xi_{2}=0$ when $q=1$; while $\xi_{1}=\xi_{2}=1$ when $q\rightarrow \infty$. Thereby, we reach at
\begin{align}
\nonumber
F_{\gamma_{i}}(x)&\approx \frac{1}{(N-M+1)!}\\
&\times \left(\frac{x/p}{\beta_{\rm UB}+\beta_{\rm LB}\beta_{\rm UL} L\left(1-\frac{8 \alpha^{4}}{\pi^{2}}\right)}\right)^{N-M+1}, L\rightarrow \infty,q=1,
\label{CDFSNRasyq1}
\end{align}
and
\begin{align}
\nonumber
F_{\gamma_{i}}(x)&\approx \frac{1}{(N-M+1)!}\\
&\times \left(\frac{x/p}{\beta_{\rm UB}+\beta_{\rm LB}\beta_{\rm UL} 2 L\left(1-\alpha^{4}\right)}\right)^{N-M+1}, \quad \{L,q\}\rightarrow \infty.
\label{CDFSNRasyqLarge}
\end{align}
{\color{black}To verify the accuracy and tightness of the proposed asymptotic approximations, Fig.~\ref{fig22} illustrates \eqref{CDFSNR} and \eqref{CDFSNRasyq1} as well as \eqref{CDFSNRasyqLarge} for the cases when $q=1$ and $q=10$ (i.e., large $q$), correspondingly. The asymptotic approach gets tighter for increasing $L$ which is in agreement to the above analysis. 
\begin{figure}[!t]
\centering
\includegraphics[trim=1.5cm .5cm .5cm 0.0cm, clip=true,totalheight=0.4\textheight]{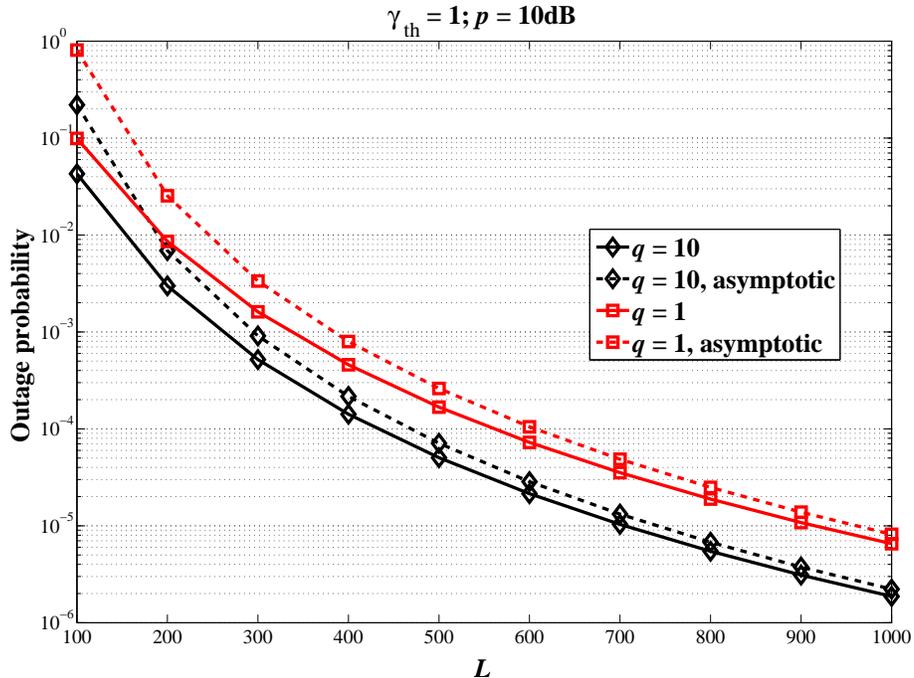}
\caption{Asymptotic and non-asymptotic CDF of the received SNR (i.e., outage probability) for different quantization bit levels $q$ vs. various values of the LRIS array size $L$. Without loss of generality and for ease of clarity, we set $\{\beta_{\rm UB}=10^{-4},\beta_{\rm LB}=10^{-2},\beta_{\rm UL}=10^{-1}\}$, $M=4$, $N=8$, $p=10$dB, and $\kappa_{\mathbf{H}}=\kappa_{\mathbf{G}}\triangleq 3$dB. Outage threshold is defined as $\gamma_{\rm th}\triangleq 2^{R}-1$ where $R=1$bps/Hz is the target (normalized) data rate.}
\label{fig22}
\end{figure}}

Moreover, when the signal of all the involved links undergo Rayleigh channel fading, i.e., $\kappa_{\mathbf{H}}=\kappa_{\mathbf{G}}=0$ within \eqref{alphapar}, then $\alpha=\sqrt{\pi}/2$. On the contrary, when the LRIS-involved links undergo pure LoS signal propagation (i.e., when $\{\kappa_{\mathbf{H}},\kappa_{\mathbf{G}}\}\rightarrow \infty$), it can be easily shown that \eqref{alphapar} becomes $\alpha=1$ in the limit. A number of useful engineering insights can be drawn from \eqref{CDFSNRasy} and the latter observations; they are summarized as follows:
\begin{itemize}
	\item For the worst-case LRIS when $L\rightarrow \infty$ yet with $q=1$ quantization bit (i.e., low-cost large-scale passive elements), Rayleigh channel fading is more impactful than pure LoS signal propagation. This occurs due to the presence of a rich scattering environment of the former rather than the latter case.\footnote{This is a well-known result for conventional MIMO systems \cite{j:JinShi07}; we show that it still holds for LRIS-enabled MIMO systems.} Specifically, 
\begin{align*}	
F_{\gamma_{i}}(x)\propto (\beta_{\rm UB}+\beta_{\rm LB}\beta_{\rm UL} L/2)^{-(N-M+1)}
\end{align*}
for Rayleigh fading, while 
\begin{align*}
F_{\gamma_{i}}(x)\propto [\beta_{\rm UB}+\beta_{\rm LB}\beta_{\rm UL} L(1-8/\pi^{2})]^{-(N-M+1)}
\end{align*}
for pure LoS signal propagation. 
	\item For the best-case LRIS scenario when $\{L,q\}\rightarrow \infty$ (i.e., high-cost), 
\begin{align*}
F_{\gamma_{i}}(x)\propto [\beta_{\rm UB}+\beta_{\rm LB}\beta_{\rm UL} 2 L (1-\pi^{2}/16)]^{-(N-M+1)}
\end{align*}
for Rayleigh fading, while 
\begin{align*}
F_{\gamma_{i}}(x)\propto \beta_{\rm UB}^{-(N-M+1)}
\end{align*}
for pure LoS signal propagation (which implies that the volume of $L$ does not influence the system performance in the absence of both channel fading and phase errors - yet, quite infeasible in practice). 
	\item When the LRIS controller has a complete lack of CSI knowledge (e.g., applying random or fixed phase shifts), then $\xi_{1}=\xi_{2}=0$ since the phase mismatches at LRIS are uniformly distributed in $[-\pi,\pi]$. Then, the SNR CDF reduces to 
\begin{align*}
F_{\gamma_{i}}(x)\propto (\beta_{\rm UB}+\beta_{\rm LB}\beta_{\rm UL}/L)^{-(N-M+1)} \quad L\rightarrow \infty. 
\end{align*}
\end{itemize}

\subsection{Imperfect CSI}
In practice, CSI estimation can be achieved via pilot-signal transmission. However, pilot signaling may be prohibitively demanding in LRIS-enabled MIMO systems because of the extremely large channel dimension regarding the transmitter-to-LRIS (i.e., $M\times L$) and receiver-to-LRIS (i.e., $L\times N$) links. A viable solution arises if the receiver acquires knowledge of only the aggregated end-to-end channel (say, $\mathbf{H G}$) and not of the individual ones; therefore, the channel dimension still remains $N\times M$, as in conventional MIMO systems. For the considered block fading case, each transmission frame is divided into two consecutive phases; namely, the training and data transmission \cite{j:NikMiridakis2017TVT,j:MiridakisTsiftsis2017}. At the former phase, $M$ pilot signals are transmitted for a duration of $T_{\textrm{pilot}}\geq M (L+1)$ time-sample instances (i.e., one corresponds to the direct link). In what follows, we assume that $T_{\textrm{pilot}}=M (L+1)$ and the transmit SNR in the training phase is the same as in the data phase, i.e., $p$. In fact, the common approach followed here is to turn off\footnote{The state `\emph{turn off}' at each LRIS element requires a special treatment in order not to act as a typical scatterer. A practical implementation of this condition is presented in \cite{j:FImani2020}.} LRIS for the first $M$ time instances where the receiver acquires CSI for the direct channel link; then, for the subsequent $L M$ time instances, the LRIS controller turns on each passive element one-by-one \cite{c:MishraJohansson2019} so as to acquire the associated cascaded channel.\footnote{The reason why the total cascaded channel appears as a superposition of $L$ one-rank matrices will be manifested in the next section. It turns out that the said formation is a requisite for phase shift optimization at LRIS.} Afterwards, the equalization ZF filter is being formed based on the estimated CSI and the system enters the data phase. Nonetheless, the channel estimates may not be entirely accurate reflecting on a channel estimation error. 

For a sufficiently large number of passive LRIS elements $L$, \eqref{wishart} converges almost surely, which in turn implies that the actual channel $\mathbf{H}_{\rm tot}$ and its channel estimate $\hat{\mathbf{H}}_{\rm tot}$ are jointly Gaussian. The estimated channel $\hat{\mathbf{H}}_{\rm tot}$ can be computed via the popular maximum-likelihood (or least-squares) method, as in, e.g., \cite{j:WangMurch2007,j:MiridakisTsiftsisTVT2017}. Then, assuming that the mean channel in \eqref{meanchannel} is perfectly known, the conditional expectation and covariance matrix of $\mathbf{H}_{\rm tot}$ are given, respectively, by \cite[Eqs. (15.61)-(15.62)]{b:kay1993fundamentals}, \cite{j:SiriteanuMiyanaga2012}
\begin{align}
\nonumber
\mathbb{E}[\mathbf{H}_{\rm tot}|\hat{\mathbf{H}}_{\rm tot}]&=\mathbf{X}+\left(\hat{\mathbf{H}}_{\rm tot}-\mathbf{X}\right)\mathbf{R}^{-1}_{\hat{\mathbf{H}}_{\rm tot}}\mathbf{R}_{\mathbf{H}_{\rm tot},\hat{\mathbf{H}}_{\rm tot}}\\
\nonumber
&=\mathbf{X}+\left(\hat{\mathbf{H}}_{\rm tot}-\mathbf{X}\right)\\
&\ \ \ \times \left(\frac{\beta_{\rm UB}+\frac{\bar{\gamma}\beta_{\rm LB}\beta_{\rm UL}}{K+1}}{\beta_{\rm UB}+\frac{\bar{\gamma}\beta_{\rm LB}\beta_{\rm UL}}{K+1}+\frac{1}{M p}}\right)\mathbf{I}_{M},
\label{expValue}
\end{align}
and
\begin{align}
\nonumber
\mathbf{R}_{\rm e}&=\mathbb{E}\bigg[\left(\mathbf{H}_{\rm tot}-\mathbb{E}[\mathbf{H}_{\rm tot}|\hat{\mathbf{H}}_{\rm tot}]\right)\\
\nonumber
&\ \ \ \times \left(\mathbf{H}_{\rm tot}-\mathbb{E}[\mathbf{H}_{\rm tot}|\hat{\mathbf{H}}_{\rm tot}]\right)^{\mathcal{H}}\big|\hat{\mathbf{H}}_{\rm tot}\bigg]\\
\nonumber
&=\mathbf{R}_{\mathbf{H}_{\rm tot}}-\mathbf{R}_{\mathbf{H}_{\rm tot},\hat{\mathbf{H}}_{\rm tot}}\mathbf{R}^{-1}_{\hat{\mathbf{H}}_{\rm tot}}\mathbf{R}^{\mathcal{H}}_{\mathbf{H}_{\rm tot},\hat{\mathbf{H}}_{\rm tot}}\\
&=\left(\frac{\beta_{\rm UB}+\frac{\bar{\gamma}\beta_{\rm LB}\beta_{\rm UL}}{K+1}}{M p\left(\beta_{\rm UB}+\frac{\bar{\gamma}\beta_{\rm LB}\beta_{\rm UL}}{K+1}\right)+1}\right)\mathbf{I}_{M},
\label{Cov}
\end{align} 
{\color{black}where 
\begin{align*}
\mathbf{R}_{\mathbf{H}_{\rm tot}}&=\mathbb{E}\left[\left(\mathbf{H}_{\rm tot}-\mathbf{X}\right)\left(\mathbf{H}_{\rm tot}-\mathbf{X}\right)^{\mathcal{H}}\right]\\
&=\left(\beta_{\rm UB}+\frac{\bar{\gamma}\beta_{\rm LB}\beta_{\rm UL}}{K+1}\right)\mathbf{I}_{M}, 
\end{align*}
\begin{align*}
\mathbf{R}_{\hat{\mathbf{H}}_{\rm tot}}&=\mathbb{E}\left[\left(\hat{\mathbf{H}}_{\rm tot}-\mathbf{X}\right)\left(\hat{\mathbf{H}}_{\rm tot}-\mathbf{X}\right)^{\mathcal{H}}\right]\\
&=\left(\beta_{\rm UB}+\frac{\bar{\gamma}\beta_{\rm LB}\beta_{\rm UL}}{K+1}+\frac{1}{M p}\right)\mathbf{I}_{M},
\end{align*}
and
\begin{align*}
\mathbf{R}_{\mathbf{H}_{\rm tot},\hat{\mathbf{H}}_{\rm tot}}&=\mathbb{E}\left[\left(\mathbf{H}_{\rm tot}-\mathbf{X}\right)\left(\hat{\mathbf{H}}_{\rm tot}-\mathbf{X}\right)^{\mathcal{H}}\right]\\
&=\left(\beta_{\rm UB}+\frac{\bar{\gamma}\beta_{\rm LB}\beta_{\rm UL}}{K+1}\right)\mathbf{I}_{M},
\end{align*}
represent the auto-covariance matrix between each column of $\mathbf{H}_{\rm tot}$, auto-covariance matrix between each column of $\hat{\mathbf{H}}_{\rm tot}$ and cross-covariance matrix between each column of $\mathbf{H}_{\rm tot}$ with $\hat{\mathbf{H}}_{\rm tot}$, respectively.}\footnote{{\color{black}In essence, for the considered least-squares method, $T_{\rm s}\geq M$ training symbols are required; thereby, $\mathbf{R}_{\hat{\mathbf{H}}_{\rm tot}}=(\beta_{\rm UB}+\frac{\bar{\gamma}\beta_{\rm LB}\beta_{\rm UL}}{K+1}+\frac{1}{T_{\rm s} p})\mathbf{I}_{M}$ \cite{j:SiriteanuMiyanaga2012}. Herein, we set $T_{\rm s}=M$.}} In the sequel, for notation simplicity, let $\overline{\mathbf{H}}_{\rm tot}\triangleq \mathbb{E}[\mathbf{H}_{\rm tot}|\hat{\mathbf{H}}_{\rm tot}]$. It is worthy to stress that $\mathbf{R}^{-1}_{\hat{\mathbf{H}}_{\rm tot}}\mathbf{R}_{\mathbf{H}_{\rm tot},\hat{\mathbf{H}}_{\rm tot}}\rightarrow \mathbf{I}_{N}$ when CSI is accurate; thus $\overline{\mathbf{H}}_{\rm tot}\rightarrow \hat{\mathbf{H}}_{\rm tot}$. On the other hand, $\mathbf{R}^{-1}_{\hat{\mathbf{H}}_{\rm tot}}\mathbf{R}_{\mathbf{H}_{\rm tot},\hat{\mathbf{H}}_{\rm tot}}\rightarrow \mathbf{0}_{N}$ when CSI is poor; which returns $\overline{\mathbf{H}}_{\rm tot}\rightarrow \mathbf{X}$. Otherwise, there is a CSI-dependent combination between the aforementioned two extremes cases \cite{j:SiriteanuMiyanaga2012}. It follows that
\begin{align}
\mathbf{H}_{\rm tot}=\overline{\mathbf{H}}_{\rm tot}+\mathbf{E},
\label{channelestimError}
\end{align} 
where $\mathbf{E}\overset{\text{d}}=\mathcal{CN}(\mathbf{0},\mathbf{R}_{\rm e})$ is the channel estimation error matrix which is uncorrelated with $\overline{\mathbf{H}}_{\rm tot}$.

Substituting \eqref{channelestimError} into \eqref{received} yields
\begin{align}
\mathbf{y}=\sqrt{p} \overline{\mathbf{H}}_{\rm tot} \mathbf{s}+\sqrt{p} \mathbf{E}\mathbf{s}+\mathbf{n},
\label{receivedEstimated}
\end{align}
where $\sqrt{p} \overline{\mathbf{H}}_{\rm tot}$ stands for the effective channel gain and $\sqrt{p} \mathbf{E}\mathbf{s}+\mathbf{n}$ is the colored-noise signal vector caused by the superposition of channel estimation errors and AWGN, which has zero-mean with equal variance per stream 
\begin{align}
\nonumber
&\mathcal{V}\triangleq \mathbb{E}\left[(\sqrt{p} \mathbf{E}\mathbf{s}+\mathbf{n})^{\mathcal{H}}(\sqrt{p} \mathbf{E}\mathbf{s}+\mathbf{n})\right]\\
&=p\rm{Tr}[\mathbf{R}_{\rm e}]+1=\frac{\beta_{\rm UB}+\frac{\bar{\gamma}\beta_{\rm LB}\beta_{\rm UL}}{K+1}}{\beta_{\rm UB}+\frac{\bar{\gamma}\beta_{\rm LB}\beta_{\rm UL}}{K+1}+1}+1.
\end{align}
Consequently, the ZF equalization filter and the post-processed received signal become, respectively, in this case \cite{j:SiriteanuMiyanaga2012} $\hat{\mathbf{W}}\triangleq \overline{\mathbf{H}}_{\rm tot}^{\dagger}$ and $\hat{\mathbf{r}}=\hat{\mathbf{W}} \mathbf{y}$. In a similar basis as in deriving \eqref{snr}, we get
\begin{align}
\gamma'_{i}=\frac{p/\mathcal{V}}{\left[\left(\overline{\mathbf{H}}_{\rm tot}^{\mathcal{H}}\overline{\mathbf{H}}_{\rm tot}\right)^{-1}\right]_{i,i}},
\label{snrEst}
\end{align}
where $\gamma'_{i}$ denotes the SNR of the $i^{\rm th}$ stream ($1\leq i\leq M$) in the presence of CSI estimation error. It follows that 
\begin{align}
\overline{\mathbf{H}}_{\rm tot}^{\mathcal{H}}\overline{\mathbf{H}}_{\rm tot}\approx \mathcal{W}_{M}(N;\hat{\mathbf{\Sigma}}),
\label{wishartEst}
\end{align}
where
\begin{align}
\hat{\mathbf{\Sigma}}\triangleq \underbrace{\left(\frac{\left(\beta_{\rm UB}+\frac{\bar{\gamma}\beta_{\rm LB}\beta_{\rm UL}}{K+1}\right)^{2}}{\beta_{\rm UB}+\frac{\bar{\gamma}\beta_{\rm LB}\beta_{\rm UL}}{K+1}+\frac{1}{M p}}\right)}_{\triangleq \mathcal{\delta}}\mathbf{I}_{M}+\frac{K \beta_{\rm LB}\beta_{\rm UL}}{N (K+1)}\mathbf{X}^{\mathcal{H}}\mathbf{X}.
\label{sigmaEst}
\end{align}
With the aid the Sherman-Morrison formula and following a similar strategy as in \eqref{sigmainverse}, it turns out that
\begin{align}
\nonumber
[\hat{\mathbf{\Sigma}}^{-1}]_{i,i}&=\frac{1}{\mathcal{\delta}}-\frac{K^{2}\beta_{\rm LB}\beta_{\rm UL}}{\left((K+1)\mathcal{\delta}\right)^{2}}\\
&\times \frac{1}{\left(1+\frac{\beta_{\rm LB}\beta_{\rm UL}K^{2} \left(1+K+\beta_{\rm UB} M p+\beta_{\rm LB}\beta_{\rm UL}\bar{\gamma} M p+\beta_{\rm UB} K M p\right)}{(1+K)(\beta_{\rm UB}+\beta_{\rm LB}\beta_{\rm UL}\bar{\gamma}+\beta_{\rm UB} K)^{2}p}\right)}.
\label{sigmainverseEst}
\end{align}
Note that $\hat{\mathbf{\Sigma}}\rightarrow \mathbf{\Sigma}$ as $p\rightarrow \infty$ reflecting that the channel estimation error tends to zero for an increasing transmit SNR. The distribution of $\gamma'_{i}$ is thus approached by \eqref{CDFSNR}, by substituting $p$ and $[\mathbf{\Sigma}^{-1}]_{i,i}$ with $p/\mathcal{V}$ and $[\hat{\mathbf{\Sigma}}^{-1}]_{i,i}$, correspondingly.

\section{Phase Shift Design at LRIS}
For the considered spatial multiplexing mode of operation, we aim to maximize the minimum received SNR, i.e., $\max \gamma_{\min}$, where $\gamma_{\min}\triangleq \min \{\gamma_{i}\}^{M}_{i=1}$. In addition, we formulate the corresponding optimization problem by assuming first an ideal phase shift setting and we provide a discrete-phase adjustment afterwards.\footnote{It was recently proven in \cite{j:XuRenzo2021} that the performance loss between the ideal and discrete phase shifts is negligible at high SNR regions when $q>2$.}

\subsection{Instantaneous CSI}
The perfect CSI case is examined first, followed by the imperfect CSI afterwards. Using similar lines of reasoning as in \cite{j:ZhangShuowen2020} and \cite{j:KimSucheol2022}, it is convenient for the subsequent analysis to reformulate $\mathbf{H}_{\rm tot}$ as 
\begin{align}
\mathbf{H}_{\rm tot}\triangleq \mathbf{H}_{-i}+e^{j \phi_{i}}\mathbf{h}_{i}\mathbf{g}^{\mathcal{H}}_{i},
\end{align}
where $\mathbf{H}_{-i}\triangleq \mathbf{H}_{\rm D}+\sum^{L}_{l=1,l\neq i}e^{j \phi_{l}}\mathbf{h}_{l}\mathbf{g}^{\mathcal{H}}_{l}$, while $\mathbf{h}_{k}\triangleq [\mathbf{H}]_{:,k}$ and $\mathbf{g}^{\mathcal{H}}_{k}\triangleq [\mathbf{G}]_{k,:}$. Based on the latter formulation, $\mathbf{H}_{\rm tot}$ is converted into a full-rank $\mathbf{H}_{\rm D}$ plus an $L-$sum series of rank-one matrices, whereas each rank-one matrix depends on the phase of each corresponding LRIS element. Hence, maximizing $\gamma_{i}$ is equivalent to maximize $\det[\mathbf{H}^{\mathcal{H}}_{\rm tot}\mathbf{H}_{\rm tot}]$ in the nominator of \eqref{snr}. 

We commence by introducing the auxiliary notations: 
\begin{align}
\nonumber
&\mathbf{A}_{i}\triangleq \mathbf{H}^{\mathcal{H}}_{-i}\mathbf{H}_{-i}+\mathbf{g}_{i}\mathbf{h}^{\mathcal{H}}_{i}\left(\mathbf{g}_{i}\mathbf{h}^{\mathcal{H}}_{i}\right)^{\mathcal{H}}; \mathbf{p}_{i}\triangleq \mathbf{g}_{i}; \mathbf{q}_{i}\triangleq \mathbf{H}^{\mathcal{H}}_{-i} \mathbf{h}_{i};\\
&\textrm{and }t_{i}\triangleq e^{j \phi_{i}}.
\end{align}
Then, after direct manipulations, the introduced optimization problem becomes
\begin{align}
&\max_{\phi_{i}}\: \det\left[\mathbf{H}^{\mathcal{H}}_{\rm tot}\mathbf{H}_{\rm tot}\right]\\
\nonumber
&\textrm{subject to }\:0\leq \phi_{i}<2\pi.
\end{align}
It turns out that
\begin{align}
\nonumber
\max_{\phi_{i}}\: \det\left[\mathbf{H}^{\mathcal{H}}_{\rm tot}\mathbf{H}_{\rm tot}\right]&= \max_{\phi_{i}}\: \det\left[\mathbf{A}_{i}+t^{\star}_{i}\mathbf{p}_{i}\mathbf{q}^{\mathcal{H}}_{i}+t_{i}\mathbf{q}_{i}\mathbf{p}^{\mathcal{H}}_{i}\right]\\
\nonumber
&= \max_{\phi_{i}}\: \det\left[\mathbf{A}_{i}+\left[\mathbf{p}_{i},\mathbf{q}_{i}\right]\diag\{[t^{\star}_{i},t_{i}]^{\mathcal{T}}\}\left[\mathbf{q}_{i},\mathbf{p}_{i}\right]^{\mathcal{H}}\right]\\
\nonumber
&= \max_{\phi_{i}}\: \det\left[\diag\{[1/t^{\star}_{i},1/t_{i}]^{\mathcal{T}}\}+\left[\mathbf{q}_{i},\mathbf{p}_{i}\right]^{\mathcal{H}}\mathbf{A}^{-1}_{i}\left[\mathbf{p}_{i},\mathbf{q}_{i}\right]\right]\\
&\times \det\left[\diag\{[t^{\star}_{i},t_{i}]^{\mathcal{T}}\}\right]\det\left[\mathbf{A}_{i}\right],
\label{optProblem}
\end{align} 
where the last equality of \eqref{optProblem} is derived by utilizing the matrix determinant Lemma. Note that $\det[\diag\{[t^{\star}_{i},t_{i}]^{\mathcal{T}}\}]=1$ and $\det[\mathbf{A}_{i}]$ is independent of $\phi_{i}$. Also, $\mathbf{A}_{i}$ is an $M\times M$ full-rank matrix (due to the implicit presence of the positive definite matrix $\mathbf{H}_{\rm D}$) and thus is invertible. It turns out that the optimal $\phi_{i}$ reads as
\begin{align}
\nonumber
\phi^{\star}_{i}&=\argmax_{\phi_{i}}\: \det\left[\diag\left\{\left[\frac{1}{t^{\star}_{i}},\frac{1}{t_{i}}\right]^{\mathcal{T}}\right\}+\left[\mathbf{q}_{i},\mathbf{p}_{i}\right]^{\mathcal{H}}\mathbf{A}^{-1}_{i}\left[\mathbf{p}_{i},\mathbf{q}_{i}\right]\right]\\
\nonumber
&=\argmax_{\phi_{i}}\: {\rm Re}\left\{e^{j \phi_{i}}\mathbf{p}^{\mathcal{H}}_{i}\mathbf{A}^{-1}_{i}\mathbf{q}_{i}\right\}=-\angle \left[\mathbf{p}^{\mathcal{H}}_{i}\mathbf{A}^{-1}_{i}\mathbf{q}_{i}\right]\\
&=-\angle \left[\mathbf{g}^{\mathcal{H}}_{i}\left(\mathbf{H}^{\mathcal{H}}_{-i}\mathbf{H}_{-i}+\mathbf{g}_{i}\mathbf{h}^{\mathcal{H}}_{i}\left(\mathbf{g}_{i}\mathbf{h}^{\mathcal{H}}_{i}\right)^{\mathcal{H}}\right)^{-1}\mathbf{H}^{\mathcal{H}}_{-i} \mathbf{h}_{i}\right].
\label{optPhi}
\end{align} 
However, since the receiver has CSI knowledge only for the cascaded channel $\mathbf{h}_{i}\mathbf{g}^{\mathcal{H}}_{i}$ (and not for the individual channels $\mathbf{h}_{k}$ and $\mathbf{g}_{k}$), the optimal $\phi_{i}$ can be computed as
\begin{align}
\nonumber
&\phi^{\star}_{i}= -\angle \:{\rm Tr}\left[\mathbf{p}^{\mathcal{H}}_{i}\mathbf{A}^{-1}_{i}\mathbf{q}_{i}\right]\\
&= -\angle \:{\rm Tr}\left[\mathbf{H}^{\mathcal{H}}_{-i}\left(\mathbf{h}_{i}\mathbf{g}^{\mathcal{H}}_{i}\right)\left(\mathbf{H}^{\mathcal{H}}_{-i}\mathbf{H}_{-i}+\left(\mathbf{h}_{i}\mathbf{g}^{\mathcal{H}}_{i}\right)^{\mathcal{H}}\left(\mathbf{h}_{i}\mathbf{g}^{\mathcal{H}}_{i}\right)\right)^{-1}\right].
\label{optPhiTrace}
\end{align}  
The latter expression implies an infinite range of phase shifts at LRIS elements. In practice, $\phi^{\star}_{i}$ becomes the closest value between \eqref{optPhiTrace} and all the available phase shifts within $|\mathcal{S}|$. For the realistic condition with estimated thus imperfect CSI, the aforementioned process is identical by substituting $\mathbf{H}_{\rm tot}$ with $\overline{\mathbf{H}}_{\rm tot}$ as per \eqref{expValue}.

Algorithm~1 summarizes the proposed phase shift design. Regarding the computational complexity of Algorithm~1 and retaining our focus on the matrix inverse and matrix multiplication operations, there is a complexity of $\mathcal{O}\{L (3M^{2}N+2 M^{3})\}$, which is linear on the number of LRIS elements and receive antennas.

\begin{algorithm}[t]
	\caption{Phase Shift Design for Instantaneous CSI}
	\begin{algorithmic}[1]
		\INPUT{$|\mathcal{S}|$; $\mathbf{H}_{\rm tot}=\mathbf{H}_{\rm D}+\sum^{L}_{l=1}e^{j \phi_{l}}\mathbf{h}_{l}\mathbf{g}^{\mathcal{H}}_{l}$; arbitrary $\{\phi_{l}\}^{L}_{l=1}$}
		\STATE{Compute \eqref{snr} and find minimum SNR $\gamma_{\min}$;}
		 \WHILE{$\gamma_{\min}<\gamma_{\min}+\epsilon$ ($\epsilon>0$ is some tolerance)}
			\FOR{$l=1:L$}
			\STATE{Update $\phi_{l}$ as $\phi_{l}=\min\{|\textrm{Eq.}~\eqref{optPhiTrace}-\mathcal{S}_{v}|\}^{|\mathcal{S}|}_{v=1}$;}
			\ENDFOR 
			\STATE{Compute SNR of $i^{\rm th}$ stream as per \eqref{snr};}
			\STATE{The new minimum SNR is $\gamma^{({\rm new})}_{\min}$;}
				\IF{$\gamma^{({\rm new})}_{\min}<\gamma_{\min}+\epsilon$}
					\STATE Break;
				\ELSE
					\STATE {$\gamma^{({\rm new})}_{\min}=\gamma_{\min}$};
				\ENDIF		   
		\ENDWHILE 
		\OUTPUT{$\{\phi^{\star}_{l}\}^{L}_{l=1}$}
	\end{algorithmic}
\end{algorithm}

\subsection{Statistical CSI}
In this case, the LRIS controller adapts the $L$ phase shifts according to a statistical-only knowledge of the actual (instantaneous) CSI. This strategy may be more feasible (or even more desirable in certain cases) than its instantaneous CSI counterpart, due to its considerably lower computational cost. In what follows, we use the imperfect CSI approach (as analyzed in the previous section) to meet realistic conditions; whereas the corresponding approach based on a perfect CSI acquisition is included as a special case. 

According to Jensen's inequality and with the aid of \cite[Thm.~3.2.12]{b:multivariate}, the average SNR of the $i^{\rm th}$ stream is lower bounded as
\begin{align}
\nonumber
\bar{\gamma}_{i}&=\mathbb{E}\left[\frac{p/\mathcal{V}}{\left[\left(\overline{\mathbf{H}}^{\mathcal{H}}_{\rm tot}\overline{\mathbf{H}}_{\rm tot}\right)^{-1}\right]_{i,i}}\right]\\
\nonumber
&\geq \frac{p/\mathcal{V}}{\mathbb{E}\left[\left[\left(\overline{\mathbf{H}}^{\mathcal{H}}_{\rm tot}\overline{\mathbf{H}}_{\rm tot}\right)^{-1}\right]_{i,i}\right]}\\
\nonumber
&=\frac{(N-M)p/\mathcal{V}}{\left[\hat{\mathbf{\Sigma}}^{-1}\right]_{i,i}}\\
&=\left(\frac{(N-M) p}{\mathcal{V}}\right)\frac{\det[\hat{\mathbf{\Sigma}}]}{\det[\hat{\mathbf{\Sigma}}_{i}]},\quad N>M.
\label{avSNR}
\end{align} 
Thereby, maximizing $\phi_{i}$ is equivalent to maximize $\det[\hat{\mathbf{\Sigma}}]$, where $\hat{\mathbf{\Sigma}}_{i}$ is the deflated version of $\hat{\mathbf{\Sigma}}$ with its $i^{\rm th}$ column being removed.\footnote{Recall that $\delta$ in \eqref{sigmaEst} reduces to $\beta_{\rm UB}+\bar{\gamma}\beta_{\rm LB}\beta_{\rm UL}/(K+1)$ by setting $p\rightarrow \infty$. Doing so, $\hat{\mathbf{\Sigma}}$ in \eqref{sigmaEst} becomes $\mathbf{\Sigma}$ referring back to \eqref{sigma}, thus reflecting on the perfect CSI condition.}

By defining $\bm{\mu}_{{\rm LB},l}\triangleq [\mathbf{M}_{\rm LB}]_{:,l}$ and $\bm{\mu}_{{\rm UL},l}\triangleq [\mathbf{M}_{\rm UL}]^{\mathcal{H}}_{l,:}$, the matrix $\mathbf{X}$ in \eqref{sigma} becomes 
\begin{align}
\mathbf{X}=\mathbf{M}_{\rm LB}\mathbf{\Phi}\mathbf{M}_{\rm UL}=\sum^{L}_{l=1}e^{j\phi_{l}}\bm{\mu}_{{\rm LB},l} \bm{\mu}^{\mathcal{H}}_{{\rm UL},l}.
\end{align} 
Then, expanding \eqref{sigma}, the optimization problem reads as
\begin{align}
\max_{\phi_{i}}\: \det\left[\hat{\mathbf{\Sigma}}\right]= \max_{\phi_{i}}\: \det\left[\hat{\mathbf{A}}_{i}+t^{\star}_{i}\hat{\mathbf{p}}_{i}\hat{\mathbf{q}}^{\mathcal{H}}_{i}+t_{i}\hat{\mathbf{q}}_{i}\hat{\mathbf{p}}^{\mathcal{H}}_{i}\right],
\label{optProblemAverage}
\end{align} 
where 
\begin{align*}
\nonumber
\hat{\mathbf{A}}_{i}&=\mathcal{\delta}\mathbf{I}_{M}\\
&+\frac{K \beta_{\rm LB}\beta_{\rm UL}}{N (K+1)}\left(\mathbf{M}^{\mathcal{H}}_{-i}\mathbf{M}_{-i}+\bm{\mu}_{{\rm UL},i}\bm{\mu}^{\mathcal{H}}_{{\rm LB},i}\left(\bm{\mu}_{{\rm UL},i}\bm{\mu}^{\mathcal{H}}_{{\rm LB},i}\right)^{\mathcal{H}}\right),
\end{align*} 
with 
\begin{align*}
\nonumber
&\mathbf{M}_{-i}\triangleq \sum^{L}_{l=1,l\neq i}e^{j\phi_{l}}\bm{\mu}_{{\rm LB},l} \bm{\mu}^{\mathcal{H}}_{{\rm UL},l};\quad \hat{\mathbf{p}}_{i}\triangleq \bm{\mu}_{{\rm UL},i};\\
&\textrm{and } \hat{\mathbf{q}}_{i}\triangleq \mathbf{M}^{\mathcal{H}}_{-i} \bm{\mu}_{{\rm LB},i}.
\end{align*} 
Notice that \eqref{optProblemAverage} is in the form of \eqref{optProblem}, while $\hat{\mathbf{A}}_{i}$ is an $M\times M$ full-rank matrix (due to the presence of identity matrix $\mathbf{I}_{M}$) and thus is invertible. Then, in a similar basis as for the derivation of \eqref{optPhiTrace}, the optimal phase shift for the statistical CSI case is expressed as
\begin{align}
\nonumber
&\phi^{\star}_{i}= -\angle \:{\rm Tr}\left[\hat{\mathbf{p}}^{\mathcal{H}}_{i}\hat{\mathbf{A}}^{-1}_{i}\hat{\mathbf{q}}_{i}\right]\\
&= -\angle \:{\rm Tr}\Bigg[\mathbf{M}^{\mathcal{H}}_{-i}\left(\bm{\mu}_{{\rm LB},i}\bm{\mu}^{\mathcal{H}}_{{\rm UL},i}\right)\Bigg(\mathcal{\delta}\mathbf{I}_{M}+\frac{K \beta_{\rm LB}\beta_{\rm UL}}{N (K+1)}\mathbf{M}^{\mathcal{H}}_{-i}\mathbf{M}_{-i}+\left(\bm{\mu}_{{\rm LB},i}\bm{\mu}^{\mathcal{H}}_{{\rm UL},i}\right)^{\mathcal{H}}\left(\bm{\mu}_{{\rm LB},i}\bm{\mu}^{\mathcal{H}}_{{\rm UL},i}\right)\Bigg)^{-1}\Bigg].
\label{optPhiTraceAverage}
\tag{42}
\end{align}
For the realistic condition with discrete-only phase shifts, $\phi^{\star}_{i}$ becomes the closest value between \eqref{optPhiTraceAverage} and all the available phase shifts within $|\mathcal{S}|$. Algorithm~2 summarizes the proposed phase shift design which has identical complexity as Algorithm~1 (yet, with much more infrequent updates due to the slower fluctuation of the statistical CSI).

\begin{algorithm}[t]
	\caption{Phase Shift Design for Statistical CSI}
	\begin{algorithmic}[1]
		\INPUT{$|\mathcal{S}|$; $\hat{\mathbf{\Sigma}}$; arbitrary $\{\phi_{l}\}^{L}_{l=1}$}
		\STATE{Compute \eqref{avSNR} and find the lower bound of minimum average SNR $\bar{\gamma}_{\min}$;}
		 \WHILE{$\bar{\gamma}_{\min}<\bar{\gamma}_{\min}+\epsilon$}
			\FOR{$l=1:L$}
			\STATE{Update $\phi_{l}$ as $\phi_{l}=\min\{|\textrm{Eq.}~\eqref{optPhiTraceAverage}-\mathcal{S}_{v}|\}^{|\mathcal{S}|}_{v=1}$;}
			\ENDFOR 
			\STATE{Compute average SNR of $i^{\rm th}$ stream as per \eqref{avSNR};}
			\STATE{The new minimum SNR is $\bar{\gamma}^{({\rm new})}_{\min}$;}
				\IF{$\bar{\gamma}^{({\rm new})}_{\min}<\bar{\gamma}_{\min}+\epsilon$}
					\STATE Break;
				\ELSE
					\STATE {$\bar{\gamma}^{({\rm new})}_{\min}=\bar{\gamma}_{\min}$};
				\ENDIF		   
		\ENDWHILE 
		\OUTPUT{$\{\phi^{\star}_{l}\}^{L}_{l=1}$}
	\end{algorithmic}
\end{algorithm} 

\section{Numerical Results and Discussion}
In this section, the derived analytical results are verified via numerical validation (in line-curves), whereas they are cross-compared with corresponding Monte-Carlo simulations (in dot-marks). In what follows and \emph{unless} otherwise specified, $\{\beta_{\rm UB}=10^{-4},\beta_{\rm LB}=10^{-2},\beta_{\rm UL}=10^{-1}\}$ (which implies a closer receiver-to-LRIS than transmitter-to-LRIS distance, as well as a farther distance of the direct link {\color{black}\cite[Eq.~(35)]{j:NadeemQurratDebbah2020}}); $q=2$ quantization bits at LRIS; $M=4$ transmit antennas (i.e., $2\times 2$ UPA); $N=8$ receive antennas ($4\times 2$); and $L=256$ passive elements ($16\times 16$) are assumed. Also, for ease of clarity, identical Rician factors for the individual LRIS-enabled channel links are adopted, i.e., $\kappa_{\mathbf{H}}=\kappa_{\mathbf{G}}\triangleq \kappa$. Average spectral efficiency (SE) per transmitted stream is selected as the performance metric under evaluation, which is defined as
\begin{align}
\nonumber
\overline{{\rm SE}}&\triangleq \mathbb{E}[\log_{2}(1+\gamma_{i})]\\
\nonumber
&=\frac{1}{{\rm ln}(2)}\int^{\infty}_{0}{\rm ln}(1+x)f_{\gamma_{i}}(x)dx\\
\nonumber
&=\frac{1}{{\rm ln}(2)}\int^{\infty}_{0}\frac{1-F_{\gamma_{i}}(x)}{1+x}dx\\
&=\frac{\exp\left(\frac{[\mathbf{\Sigma}^{-1}]_{i,i}}{p}\right)}{{\rm ln}(2)}\sum^{N-M}_{k=0}\left(\frac{[\mathbf{\Sigma}^{-1}]_{i,i}}{p}\right)^{k}\Gamma\left(-k,\frac{[\mathbf{\Sigma}^{-1}]_{i,i}}{p}\right),
\end{align}
where the second-last equality is obtained via integration by parts and the last equality arises by utilizing \cite[Eq. (3.383.10)]{tables}.

\begin{figure}[!t]
\centering
\includegraphics[trim=3.0cm .5cm 3.0cm 0.0cm, clip=true,totalheight=0.4\textheight]{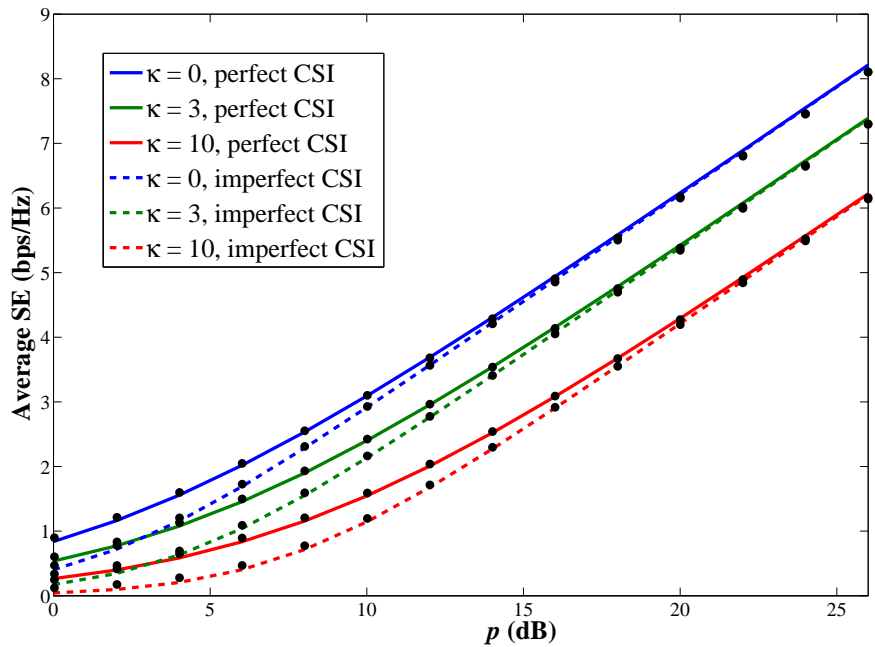}
\caption{Average SE vs. various values of the transmit SNR for the perfect and imperfect (estimated) CSI cases.}
\label{fig2}
\end{figure}

In Fig.~\ref{fig2}, the average SE is illustrated against various transmit SNR values for different CSI and channel fading conditions. Obviously, the system performance gets worse when the Rician factor of the individual LRIS-enabled links (i.e., $\kappa$) grows regardless of the CSI acquisition accuracy. This reflects on the fact that richer scattering can actually enhance the performance compared with pure LoS propagation (since rich scattering reduces the condition number of the channel matrix and hence increases the available DoF). The scenario of both perfect and imperfect/estimated CSI are compared. The arising performance difference is evident in moderately medium-to-low transmit SNR regions (e.g., less than about $15$dB), while it becomes marginal as $p$ increases.

\begin{figure}[!t]
\centering
\includegraphics[trim=2.0cm .5cm 3.0cm 0.0cm, clip=true,totalheight=0.4\textheight]{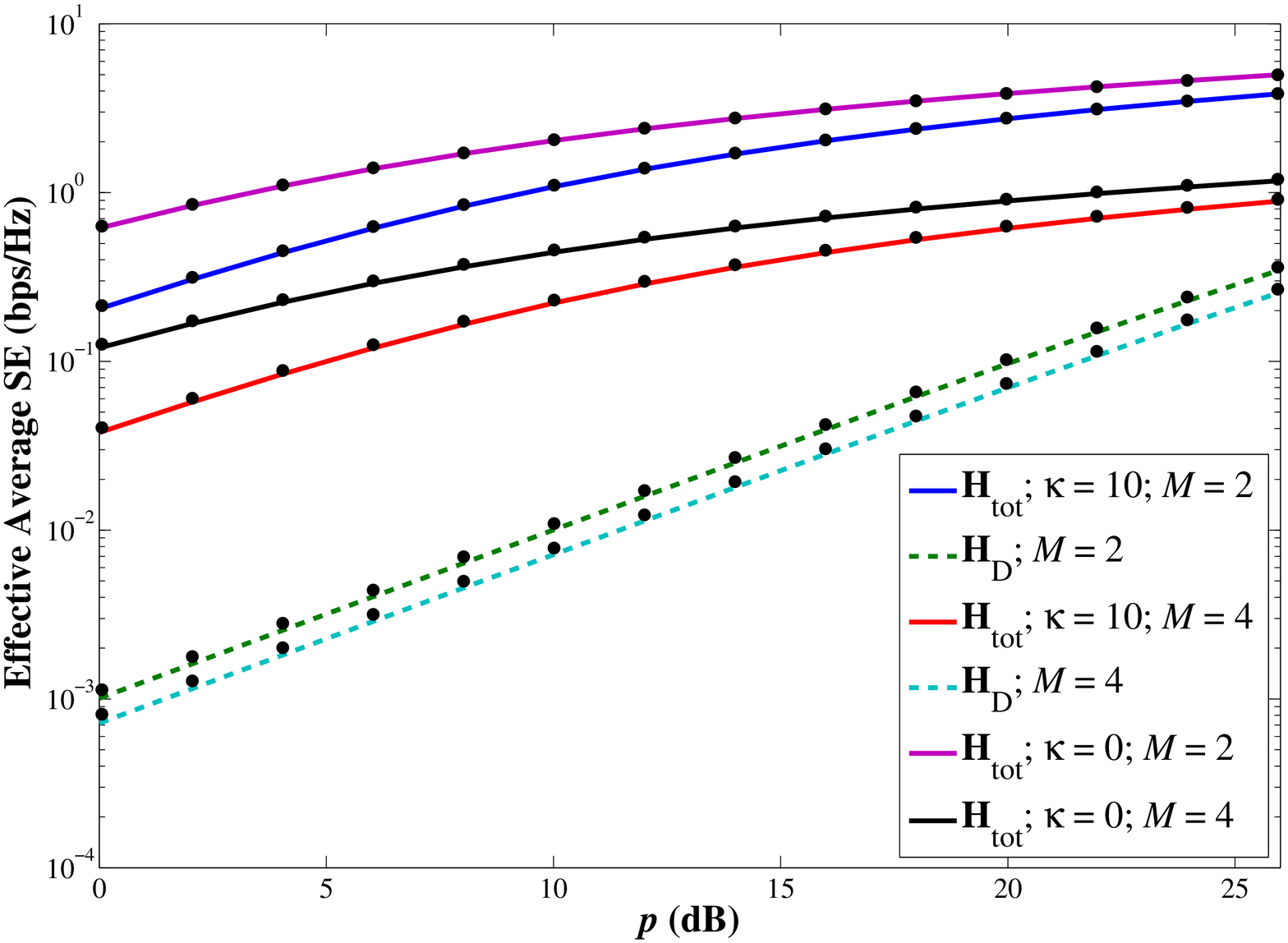}
\caption{Effective average SE vs. various values of the transmit SNR for different system configurations under a perfect CSI condition and the absence (denoted as $\mathbf{H}_{\rm D}$) or presence (denoted as $\mathbf{H}_{\rm tot}$) of LRIS, when ${\rm CT}=1200$.}
\label{fig3}
\end{figure}

In Fig.~\ref{fig3}, the signaling overhead produced by the use of LRIS is evaluated in terms of the `effective average SE' per transmitted stream, which is defined as
\begin{align}
\frac{{\rm CT}-T_{\rm pilot}}{{\rm CT}}\times \overline{{\rm SE}},
\end{align}
where ${\rm CT}$ denotes the coherence time of each transmission block (measured in consecutive time instances) and $T_{\rm pilot}=M$ or $T_{\rm pilot}=M(L+1)$ in the absence or presence of LRIS, respectively. Remarkably, the system performance benefits from the use of LRIS, despite the rather increased signaling overhead. In particular, for the considered use case of Fig.~\ref{fig3} with ${\rm CT}=1200$ (typically, ${\rm CT}$ ranges within $300$ and $10^{4}$ time instances in a vast variety of wireless communication systems) and $256$ passive elements at LRIS, the signaling overhead of acquiring $\mathbf{H}_{\rm tot}$ is about $86\%$ heavier than the case of $\mathbf{H}_{\rm D}$ alone (i.e., absence of LRIS) when $M=4$, and $43\%$ when $M=2$. Nevertheless, the said limited DoF at the LRIS case are enough to outperform the conventional communication approach based solely on the direct link; yet, this condition holds whenever a weak direct channel link $\mathbf{H}_{\rm D}$ exists, which reveals the effectiveness of LRIS in general. As expected, the system performance is being enhanced for less simultaneously transmitted independent streams (lower $M$) and/or weaker LoS links.

\begin{figure}[!t]
\centering
\includegraphics[trim=2.0cm .5cm 2.5cm 0.0cm, clip=true,totalheight=0.4\textheight]{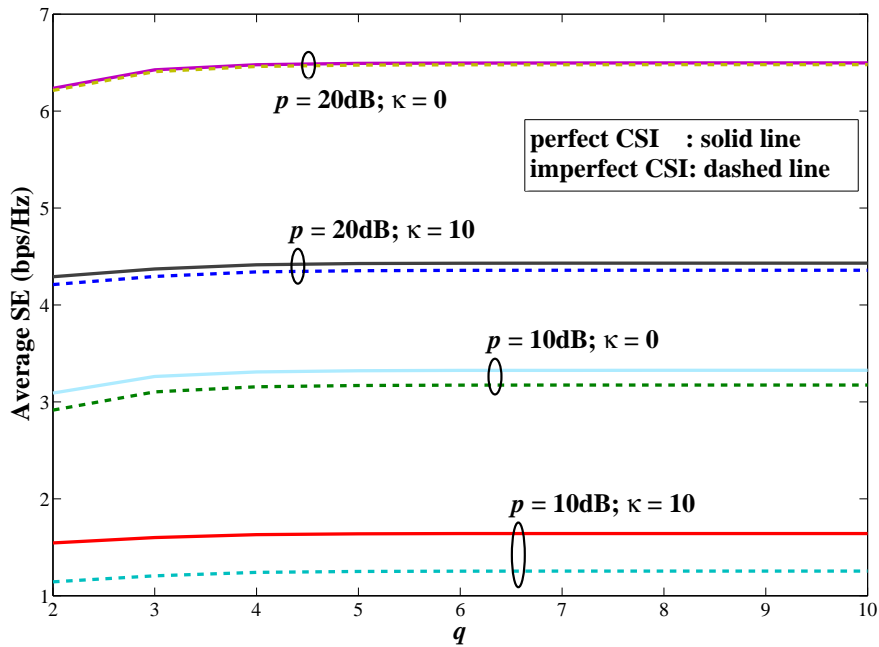}
\caption{Average SE vs. different range of phase shift quantization at LRIS, under perfect and imperfect CSI conditions.}
\label{fig4}
\end{figure}

The impact of $q-$bit quantization applied on the discrete LRIS phases is evaluated in Fig.~\ref{fig4}. Notably, the system performance rapidly converges by utilizing only a few quantization bits, whereas the performance difference tends to be marginal for $q\geq 4$. This is in accordance to \eqref{CDFSNRasyq1} and \eqref{CDFSNRasyqLarge}, where it is verified that $q-$bit quantization does not influence the overall system performance at the price of adopting a large-scale RIS. Further, it is shown that such a behavior holds for the general case, since it is independent of the transmit SNR, CSI accuracy and LoS presence/absence.

\begin{figure}[!t]
\centering
\includegraphics[trim=2.0cm .5cm 2.5cm 0.0cm, clip=true,totalheight=0.4\textheight]{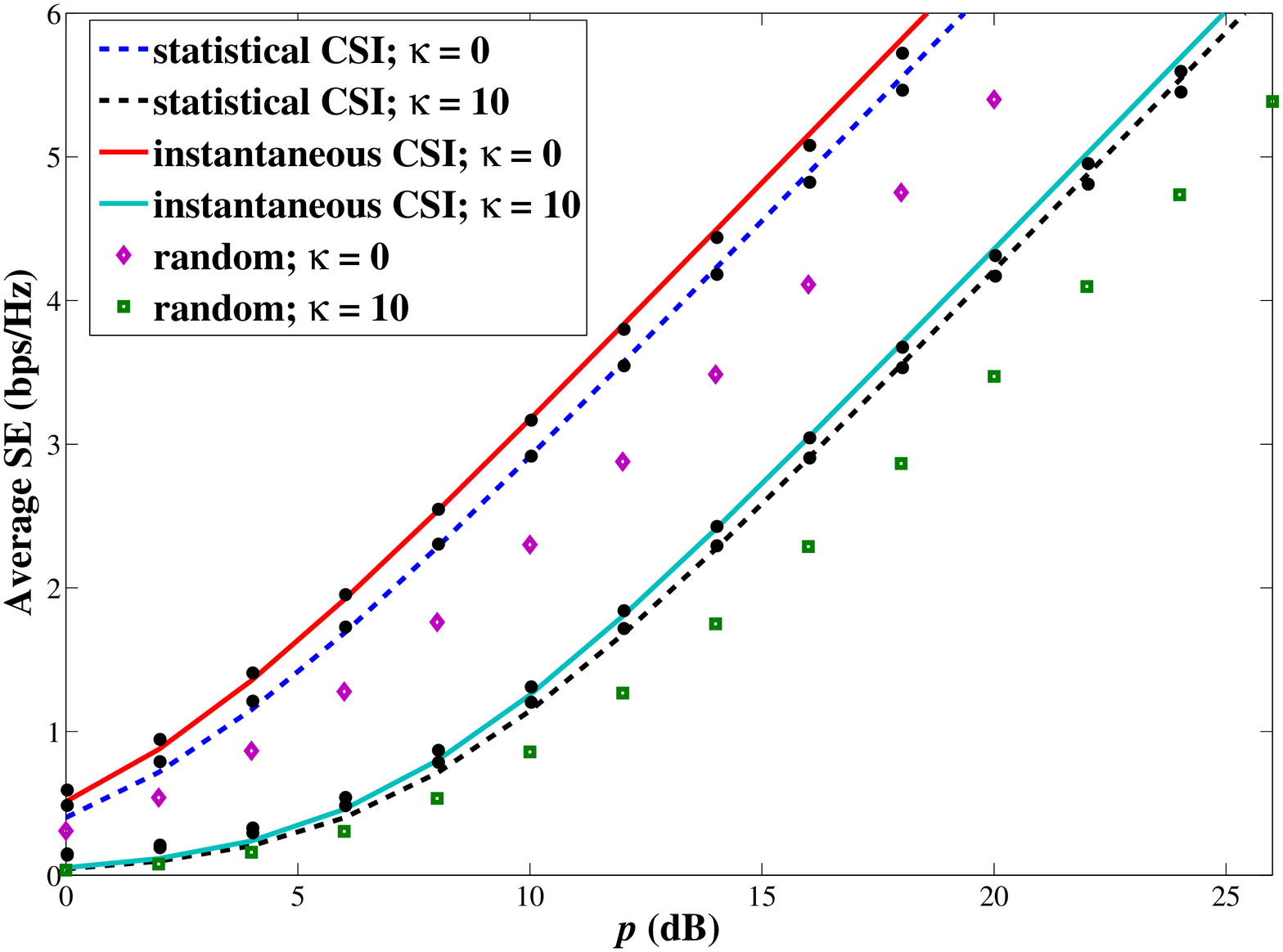}
\caption{Average SE vs. various values of the transmit SNR under imperfect (estimated) CSI and different LoS conditions.}
\label{fig5}
\end{figure}

For all the above numerical results, the phase adjustments at LRIS have been obtained with the aid of Algorithm~1 (i.e., utilizing the instantaneous CSI). In Fig.~\ref{fig5}, the discrete phase adjustments at LRIS based on the instantaneous (Algorithm~1) and statistical (Algorithm~2) CSI are compared. It is noteworthy that the two considered approaches provide a similar performance with a close gap in various SNR regimes and channel fading conditions. Again, the case of Rayleigh fading ($\kappa=0$) outperforms the one with LoS presence, as it should be. It is also observed that the aforementioned performance gap is slightly reduced in the presence of LoS signal propagation. This occurs because the instantaneous CSI tends more closely to its statistical counterpart as the available DoF of the channel matrix are being reduced (when the LoS propagation is more intense). Recall that the said condition requires a large-scale array of passive elements (and becomes tighter as $L\rightarrow \infty$); yet, using a cost-efficient phase adjustment based on the statistical CSI, which may change after several transmission block time intervals. Capitalizing on the latter observation, it turns out that phase adjustments based on statistical CSI are beneficial in large-scale RIS, since they produce an efficient performance-complexity tradeoff in contrast to the demanding instantaneous CSI counterpart. {\color{black}Finally, the case of random (discrete) phase shift rotations is also included in Fig.~\ref{fig5} as a performance benchmark. The superiority of the CSI-enabled phase adjustment at LRIS against its blind counterpart is obvious.}   

\section{Conclusions}
The MIMO uplink communication case assisted by LRIS was analytically studied, when the linear ZF detection is used at the receiver side. The practical scenario of discrete phase adjustments, Rician channel fading at the LRIS-enabled links and the existence of a direct transceiver link were considered. Two practical phase shift design schemes were presented, which are linear to the number of passive LRIS elements and receive antennas; namely, one for the instantaneous CSI and the other for statistical CSI. Both the ideal scenario of perfect CSI acquisition (performance benchmark) and estimated/imperfect CSI were analyzed. For the latter condition, asymptotic Gaussianity of the channel matrix (arising from the large-scale passive array) and its corresponding statistics were used to define an effective signal detection. Outage performance and average spectral efficiency were obtained in accurate and rather simple closed-form expressions. 

Finally, some new and useful engineering outcomes were manifested for LRIS-enabled MIMO systems, such as: (\emph{a.}) the $q-$bit quantization does not influence much the system performance; (\emph{b.}) the introduced signaling overhead using LRIS worth whenever the direct link is weak; (\emph{c.}) the presence of strong LoS propagation of MIMO signals deteriorates the overall performance; (\emph{d.}) and it is more effective (regarding the performance-complexity tradeoff) to adjust the phase shifts at LRIS according to the statistical CSI.


\bibliographystyle{IEEEtran}
\bibliography{IEEEabrv,References}

\begin{thebibliography}{10}
\providecommand{\url}[1]{#1}
\csname url@samestyle\endcsname
\providecommand{\newblock}{\relax}
\providecommand{\bibinfo}[2]{#2}
\providecommand{\BIBentrySTDinterwordspacing}{\spaceskip=0pt\relax}
\providecommand{\BIBentryALTinterwordstretchfactor}{4}
\providecommand{\BIBentryALTinterwordspacing}{\spaceskip=\fontdimen2\font plus
\BIBentryALTinterwordstretchfactor\fontdimen3\font minus
  \fontdimen4\font\relax}
\providecommand{\BIBforeignlanguage}[2]{{%
\expandafter\ifx\csname l@#1\endcsname\relax
\typeout{** WARNING: IEEEtran.bst: No hyphenation pattern has been}%
\typeout{** loaded for the language `#1'. Using the pattern for}%
\typeout{** the default language instead.}%
\else
\language=\csname l@#1\endcsname
\fi
#2}}
\providecommand{\BIBdecl}{\relax}
\BIBdecl

\bibitem{3gppR18}
\BIBentryALTinterwordspacing
{3GPP} (2022). {R}elease~18. [Online]. Available:
  \url{https://www.3gpp.org/release18}
\BIBentrySTDinterwordspacing

\bibitem{j:whitepaper2020}
``{The next hyper-connected experience for all},'' White Paper, Samsung 6G
  Vision, Jun. 2020.

\bibitem{j:TariqFaisal2020}
F.~Tariq, M.~R.~A. Khandaker, K.-K. Wong, M.~A. Imran, M.~Bennis, and
  M.~Debbah, ``A speculative study on 6{G},'' \emph{{IEEE} Wireless Commun.},
  vol.~27, no.~4, pp. 118--125, 2020.

\bibitem{j:ChengLeiDai2022}
Q.~Cheng, L.~Zhang, J.~Y. Dai, W.~Tang, J.~C. Ke, S.~Liu, J.~C. Liang, S.~Jin,
  and T.~J. Cui, ``Reconfigurable intelligent surfaces:
  {S}implified-architecture transmitters--from theory to implementations,''
  \emph{Proc. {IEEE}}, vol. 110, no.~9, pp. 1266--1289, 2022.

\bibitem{j:SwindlehurstZhou2022}
A.~L. Swindlehurst, G.~Zhou, R.~Liu, C.~Pan, and M.~Li, ``Channel estimation
  with reconfigurable intelligent surfaces--{A} general framework,''
  \emph{Proc. {IEEE}}, vol. 110, no.~9, pp. 1312--1338, 2022.

\bibitem{j:DiBoya2020}
B.~Di, H.~Zhang, L.~Song, Y.~Li, Z.~Han, and H.~V. Poor, ``Hybrid beamforming
  for reconfigurable intelligent surface based multi-user communications:
  {A}chievable rates with limited discrete phase shifts,'' \emph{{IEEE} J. Sel.
  Areas Commun.}, vol.~38, no.~8, pp. 1809--1822, 2020.

\bibitem{j:WuQingqing2020}
Q.~Wu and R.~Zhang, ``Beamforming optimization for wireless network aided by
  intelligent reflecting surface with discrete phase shifts,'' \emph{{IEEE}
  Trans. Commun.}, vol.~68, no.~3, pp. 1838--1851, 2020.

\bibitem{j:NadeemQurrat2020}
Q.-U.-A. Nadeem, H.~Alwazani, A.~Kammoun, A.~Chaaban, M.~Debbah, and M.-S.
  Alouini, ``Intelligent reflecting surface-assisted multi-user {MISO}
  communication: {C}hannel estimation and beamforming design,'' \emph{IEEE Open
  J. Commun. Soc.}, vol.~1, pp. 661--680, 2020.

\bibitem{j:ZhiKangda2021}
K.~Zhi, C.~Pan, H.~Ren, and K.~Wang, ``Statistical {CSI}-based design for
  reconfigurable intelligent surface-aided massive {MIMO} systems with direct
  links,'' \emph{{IEEE} Wireless Commun. Lett.}, vol.~10, no.~5, pp.
  1128--1132, 2021.

\bibitem{j:XiaoGechuan2022}
G.~Xiao, T.~Yang, C.~Huang, X.~Wu, H.~Feng, and B.~Hu, ``Average rate
  approximation and maximization for {RIS}-assisted multi-user {MISO} system,''
  \emph{{IEEE} Wireless Commun. Lett.}, vol.~11, no.~1, pp. 173--177, 2022.

\bibitem{j:ZhiKanda2022}
K.~Zhi, C.~Pan, H.~Ren, and K.~Wang, ``Ergodic rate analysis of reconfigurable
  intelligent surface-aided massive {MIMO} systems with {ZF} detectors,''
  \emph{{IEEE} Commun. Lett.}, vol.~26, no.~2, pp. 264--268, 2022.

\bibitem{j:WeiXiuhong2021}
X.~Wei, D.~Shen, and L.~Dai, ``Channel estimation for {RIS} assisted wireless
  communications -- {P}art {I}: {F}undamentals, solutions, and future
  opportunities,'' \emph{{IEEE} Commun. Lett.}, vol.~25, no.~5, pp. 1398--1402,
  2021.

\bibitem{j:KimSucheol2022}
S.~Kim, H.~Lee, J.~Cha, S.-J. Kim, J.~Park, and J.~Choi, ``Practical channel
  estimation and phase shift design for intelligent reflecting surface
  empowered {MIMO} systems,'' \emph{{IEEE} Trans. Wireless Commun.}, vol.~21,
  no.~8, pp. 6226--6241, 2022.

\bibitem{j:WeiLiHuangSha2022}
L.~Wei, C.~Huang, G.~C. Alexandropoulos, W.~E.~I. Sha, Z.~Zhang, M.~Debbah, and
  C.~Yuen, ``Multi-user holographic mimo surfaces: Channel modeling and
  spectral efficiency analysis,'' \emph{{IEEE} J. Sel. Topics Signal Process.},
  vol.~16, no.~5, pp. 1112--1124, 2022.

\bibitem{j:GuoYabo2022}
Y.~Guo, P.~Sun, Z.~Yuan, C.~Huang, Q.~Guo, Z.~Wang, and C.~Yuen, ``Efficient
  channel estimation for {RIS}-aided {MIMO} communications with unitary
  approximate message passing,'' \emph{{IEEE} Trans. Wireless Commun.}, to
  appear, 2022.

\bibitem{j:WeiLiHuangChongwen2022}
L.~Wei, C.~Huang, G.~C. Alexandropoulos, C.~Yuen, Z.~Zhang, and M.~Debbah,
  ``Channel estimation for {RIS}-empowered multi-user {MISO} wireless
  communications,'' \emph{{IEEE} Trans. Commun.}, vol.~69, no.~6, pp.
  4144--4157, 2021.

\bibitem{j:WeiLiHuang2022}
L.~Wei, C.~Huang, Q.~Guo, Z.~Yang, Z.~Zhang, G.~C. Alexandropoulos, M.~Debbah,
  and C.~Yuen, ``Joint channel estimation and signal recovery for
  {RIS}-empowered multiuser communications,'' \emph{{IEEE} Trans. Commun.},
  vol.~70, no.~7, pp. 4640--4655, 2022.

\bibitem{j:ZhangLiuJun2021}
J.~Zhang, J.~Liu, S.~Ma, C.-K. Wen, and S.~Jin, ``Large system achievable rate
  analysis of {RIS}-assisted {MIMO} wireless communication with statistical
  {CSIT},'' \emph{{IEEE} Trans. Wireless Commun.}, vol.~20, no.~9, pp.
  5572--5585, 2021.

\bibitem{b:multivariate}
R.~J. Muirhead, \emph{Aspects of Multivariate Statistical Theory}.\hskip 1em
  plus 0.5em minus 0.4em\relax New York: John Wiley \& Sons, 1982.

\bibitem{tables}
I.~S. Gradshteyn and I.~M. Ryzhik, \emph{Table of Integrals, Series, and
  Products}.\hskip 1em plus 0.5em minus 0.4em\relax Academic Press, 2007.

\bibitem{j:RenzoZappone20}
M.~Di~Renzo, A.~Zappone, M.~Debbah, M.-S. Alouini, C.~Yuen, J.~de~Rosny, and
  S.~Tretyakov, ``Smart radio environments empowered by reconfigurable
  intelligent surfaces: How it works, state of research, and the road ahead,''
  \emph{{IEEE} J. Sel. Areas Commun.}, vol.~38, no.~11, pp. 2450--2525, Nov.
  2020.

\bibitem{j:EmilLuca2021}
E.~Bj\"{o}rnson and L.~Sanguinetti, ``Rayleigh fading modeling and channel
  hardening for reconfigurable intelligent surfaces,'' \emph{{IEEE} Wireless
  Commun. Lett.}, vol.~10, no.~4, pp. 830--834, 2021.

\bibitem{j:WangBadiu2022}
\BIBentryALTinterwordspacing
T.~Wang, M.-A. Badiu, G.~Chen, and J.~P. Coon, ``Performance analysis of
  {IOS}-assisted {NOMA} system with channel correlation and phase errors,''
  \emph{arXiv}, 2021. [Online]. Available:
  \url{https://arxiv.org/abs/2112.11512.}
\BIBentrySTDinterwordspacing

\bibitem{j:KhiongYong2005}
S.~K. Yong and J.~Thompson, ``Three-dimensional spatial fading correlation
  models for compact {MIMO} receivers,'' \emph{{IEEE} Trans. Wireless Commun.},
  vol.~4, no.~6, pp. 2856--2869, 2005.

\bibitem{j:Badiu2020}
M.-A. Badiu and J.~P. Coon, ``Communication through a large reflecting surface
  with phase errors,'' \emph{{IEEE} Wireless Commun. Lett.}, vol.~9, no.~2, pp.
  184--188, Feb. 2020.

\bibitem{b:AlouiniSimon}
M.~K. Simon and M.-S. Alouini, \emph{Digital Communication over Fading
  Channels}.\hskip 1em plus 0.5em minus 0.4em\relax John Wiley \& Sons, Inc.,
  2005.

\bibitem{j:steyn1972approximations}
H.~S. Steyn and J.~J.~J. Roux, ``Approximations for the non-central wishart
  distribution,'' \emph{South Afr. Statist. J.}, vol.~6, no.~2, pp. 165--173,
  1972.

\bibitem{j:MatthaiouMckaySmith2010}
M.~Matthaiou, M.~R. Mckay, P.~J. Smith, and J.~A. Nossek, ``On the condition
  number distribution of complex wishart matrices,'' \emph{{IEEE} Trans.
  Commun.}, vol.~58, no.~6, pp. 1705--1717, 2010.

\bibitem{j:SiriteanuMiyanaga2012}
C.~Siriteanu, Y.~Miyanaga, S.~D. Blostein, S.~Kuriki, and X.~Shi, ``{MIMO}
  zero-forcing detection analysis for correlated and estimated {R}ician
  fading,'' \emph{{IEEE} Trans. Veh. Technol.}, vol.~61, no.~7, pp. 3087--3099,
  Sep. 2012.

\bibitem{j:JinShi07}
S.~Jin, X.~Gao, and X.~You, ``On the ergodic capacity of rank-$1$
  {R}icean-fading {MIMO} channels,'' \emph{{IEEE} Trans. Inf. Theory}, vol.~53,
  no.~2, pp. 502--517, 2007.

\bibitem{j:NikMiridakis2017TVT}
N.~I. Miridakis, T.~A. Tsiftsis, and C.~Rowell, ``Distributed spatial
  multiplexing systems with hardware impairments and imperfect channel
  estimation under {R}ank$-1$ {R}ician fading channels,'' \emph{{IEEE} Trans.
  Veh. Technol.}, vol.~66, no.~6, pp. 5122--5133, 2017.

\bibitem{j:MiridakisTsiftsis2017}
N.~I. Miridakis, T.~A. Tsiftsis, G.~C. Alexandropoulos, and M.~Debbah,
  ``Simultaneous spectrum sensing and data reception for cognitive spatial
  multiplexing distributed systems,'' \emph{{IEEE} Trans. Wireless Commun.},
  vol.~16, no.~5, pp. 3313--3327, 2017.

\bibitem{j:FImani2020}
M.~F.~Imani, D.~R. Smith, and P.~del Hougne, ``Perfect absorption in a
  disordered medium with programmable meta-atom inclusions,'' \emph{Advanced
  Functional Materials}, vol.~30, no.~52, p. 2005310, 2020.

\bibitem{c:MishraJohansson2019}
D.~Mishra and H.~Johansson, ``Channel estimation and low-complexity beamforming
  design for passive intelligent surface assisted {MISO} wireless energy
  transfer,'' in \emph{IEEE Int. Conf. Acoust., Speech and Signal Process.
  (ICASSP)}, Brighton, U.K., May 2019, pp. 4659--4663.

\bibitem{j:WangMurch2007}
C.~Wang, E.~K.~S. Au, R.~D. Murch, W.~H. Mow, R.~S. Cheng, and V.~Lau, ``On the
  performance of the {MIMO} zero-forcing receiver in the presence of channel
  estimation error,'' \emph{{IEEE} Trans. Wireless Commun.}, vol.~6, no.~3, pp.
  805--810, Mar. 2007.

\bibitem{j:MiridakisTsiftsisTVT2017}
N.~I. Miridakis and T.~A. Tsiftsis, ``On the joint impact of hardware
  impairments and imperfect {CSI} on successive decoding,'' \emph{{IEEE} Trans.
  Veh. Technol.}, vol.~66, no.~6, pp. 4810--4822, 2017.

\bibitem{b:kay1993fundamentals}
S.~M. Kay, \emph{Fundamentals of statistical signal processing: Estimation
  theory}.\hskip 1em plus 0.5em minus 0.4em\relax Prentice-Hall, Inc., 1993.

\bibitem{j:XuRenzo2021}
P.~Xu, G.~Chen, Z.~Yang, and M.~D. Renzo, ``Reconfigurable intelligent
  surfaces-assisted communications with discrete phase shifts: {H}ow many
  quantization levels are required to achieve full diversity?'' \emph{{IEEE}
  Wireless Commun. Lett.}, vol.~10, no.~2, pp. 358--362, 2021.

\bibitem{j:ZhangShuowen2020}
S.~Zhang and R.~Zhang, ``Capacity characterization for intelligent reflecting
  surface aided {MIMO} communication,'' \emph{{IEEE} J. Sel. Areas Commun.},
  vol.~38, no.~8, pp. 1823--1838, 2020.

\bibitem{j:NadeemQurratDebbah2020}
Q.-U.-A. Nadeem, A.~Kammoun, A.~Chaaban, M.~Debbah, and M.-S. Alouini,
  ``Asymptotic max-min {SINR} analysis of reconfigurable intelligent surface
  assisted {MISO} systems,'' \emph{{IEEE} Trans. Wireless Commun.}, vol.~19,
  no.~12, pp. 7748--7764, 2020.

\end{thebibliography}

\vfill

\end{document}